\documentclass[10pt,journal,compsoc]{IEEEtran}
\ifCLASSOPTIONcompsoc
  \usepackage[nocompress]{cite}
\else
  \usepackage{cite}
\fi
\usepackage{multirow}
\usepackage{graphicx}
\usepackage{amsmath}
\usepackage[usenames,dvipsnames]{color}
\usepackage[abs]{overpic}
\usepackage[T1]{fontenc}
\usepackage[utf8]{inputenc}
\usepackage{lmodern}
\usepackage{lipsum,multicol}
\usepackage{float}
\usepackage{color}
\usepackage{overpic}
\usepackage{caption}
\usepackage{subcaption}
\usepackage{multirow}
\usepackage{nameref}
\usepackage{lmodern}
\usepackage{framed,color}
\usepackage{algorithm}
\usepackage[noend]{algpseudocode}
\usepackage{algpseudocode}
\usepackage{listings}

\usepackage{soul}
\usepackage{wrapfig}
\usepackage{booktabs}
\usepackage{url}

\makeatletter
\def\BState{\State\hskip-\ALG@thistlm}
\makeatother

\makeatletter
\def\therule{\makebox[\algorithmicindent][l]{\hspace*{.5em}\vrule height .75\baselineskip depth .25\baselineskip}}%

\newtoks\therules
\therules={}
\def\appendto#1#2{\expandafter#1\expandafter{\the#1#2}}
\def\gobblefirst#1{
  #1\expandafter\expandafter\expandafter{\expandafter\@gobble\the#1}}%
    \def\LState{\State\unskip\the\therules}
    \def\pushindent{\appendto\therules\therule}%
    \def\popindent{\gobblefirst\therules}%
    \def\printindent{\unskip\the\therules}%
    \def\printandpush{\printindent\pushindent}%
    \def\popandprint{\popindent\printindent}%
    \algdef{SE}[WHILE]{While}{EndWhile}[1]
    {\printandpush\algorithmicwhile\ #1\ \algorithmicdo}
    {\popandprint\algorithmicend\ \algorithmicwhile}%
    \algdef{SE}[FOR]{For}{EndFor}[1]
    {\printandpush\algorithmicfor\ #1\ \algorithmicdo}
    {\popandprint\algorithmicend\ \algorithmicfor}%
    \algdef{S}[FOR]{ForAll}[1]
    {\printindent\algorithmicforall\ #1\ \algorithmicdo}%
    \algdef{SE}[LOOP]{Loop}{EndLoop}
    {\printandpush\algorithmicloop}
    {\popandprint\algorithmicend\ \algorithmicloop}%
    \algdef{SE}[REPEAT]{Repeat}{Until}
    {\printandpush\algorithmicrepeat}[1]
    {\popandprint\algorithmicuntil\ #1}%
    \algdef{SE}[IF]{If}{EndIf}[1]
    {\printandpush\algorithmicif\ #1\ \algorithmicthen}
    {\popandprint\algorithmicend\ \algorithmicif}%
    \algdef{C}[IF]{IF}{ElsIf}[1]
    {\popandprint\pushindent\algorithmicelse\ \algorithmicif\ #1\ \algorithmicthen}%
    \algdef{Ce}[ELSE]{IF}{Else}{EndIf}
    {\popandprint\pushindent\algorithmicelse}%
    \algdef{SE}[PROCEDURE]{Procedure}{EndProcedure}[2]
    {\printandpush\algorithmicprocedure\ \textproc{#1}\ifthenelse{\equal{#2}{}}{}{(#2)}}%
    {\popandprint\algorithmicend\ \algorithmicprocedure}%
    \algdef{SE}[FUNCTION]{Function}{EndFunction}[2]
    {\printandpush\algorithmicfunction\ \textproc{#1}\ifthenelse{\equal{#2}{}}{}{(#2)}}%
    {\popandprint\algorithmicend\ \algorithmicfunction}%
    \makeatother

\newtheorem{theorem}{Theorem}
\newtheorem{lemma}{Lemma}

\makeatletter
\newcommand\footnoteref[1]{\protected@xdef\@thefnmark{\ref{#1}}\@footnotemark}
\makeatother

\begin{document}
\title{LocalCoin: An Ad-hoc Payment Scheme for Areas with High Connectivity}
\author{Dimitris Chatzopoulos, Sujit Gujar, Boi Faltings, and Pan Hui
	\IEEEcompsocitemizethanks{
		\IEEEcompsocthanksitem Dimitris Chatzopoulos and Pan Hui are with The Hong Kong University of Science and Technology, Hong Kong. E-mail: \{dcab,panhui\}@cse.ust.hk
		\IEEEcompsocthanksitem Sujit Gujar is with the International Institute of Information Technology (IIIT), Hyderabad, India. E-mail: sujit.gujar@iiit.ac.in
		\IEEEcompsocthanksitem Boi Faltings is with the Ecole Polytechnique Federale de Lausanne (EPFL), Switzerland. E-mail: boi.faltings@epfl.ch
	}
\thanks{Manuscript received: date; revised: date}
}

\IEEEtitleabstractindextext{
\begin{abstract}
The popularity of digital currencies, especially cryptocurrencies, has been continuously growing since the appearance of Bitcoin. Bitcoin's security lies in a proof-of-work scheme, which requires high computational resources at the miners. 
Despite advances in mobile technology, existing cryptocurrencies cannot be maintained by mobile devices due to their low processing capabilities. 
Mobile devices can only accommodate mobile applications (wallets) that allow users to exchange credits of cryptocurrencies. 
In this work, we propose LocalCoin, an alternative cryptocurrency that requires minimal computational resources, produces low data traffic and works with off-the-shelf mobile devices. LocalCoin replaces the computational hardness that is at the root of Bitcoin's security with the social hardness of ensuring that all witnesses to a transaction are colluders. Localcoin features (i) a lightweight proof-of-work scheme and (ii) a distributed blockchain. We analyze LocalCoin for double spending for passive and active attacks and prove that under the assumption of sufficient number of users and properly selected tuning parameters the probability of double spending is close to zero. Extensive simulations on real mobility traces, realistic urban settings, and random geometric graphs show that the probability of success of one transaction converges to 1 and the probability of the success of a double spending attempt converges to 0.

\end{abstract}

}

\maketitle

\IEEEdisplaynontitleabstractindextext
\IEEEpeerreviewmaketitle

\IEEEraisesectionheading{\section{Introduction}\label{sec:introduction}}

\IEEEPARstart{B}{itcoin},  proposed by Nakamoto in 2009, is the most popular cryptocurrency~\cite{nakamoto2012bitcoin}.  
Numerous cryptocurrencies have been proposed thereafter and have attracted the attention of both financial and technological industries as well as academia~\cite{CoinDesk} \cite{coinmarketcap}\cite{mapofcoins}.  All the digital currencies that were proposed before Bitcoin, follow the client/server model with transactions possible only between the currency provider and the users (PAYPAL, VISA, MASTERCARD, etc). Bitcoin, in contrast, works in a decentralised manner.

Decentralised cryptocurrencies have to deal with three main challenges: \textbf{(i) Proof of ownership}- users should be able to prove they have the amount of money they claim to have. \textbf{(ii) Double spending avoidance} - a defense mechanism against double spending. (Users are not able to spend the same money more than once). \textbf{(iii) Incentives} - for its stakeholders.  Common characteristics of all the existing cryptocurrencies are: \textbf{(i)} Internet based \textbf{(ii)} use computationally expensive techniques to deal with double spending attacks and \textbf{(iii)} require lots of data storage.  To become part of the Bitcoin peer network anyone can contribute their resources and work as a \textbf{miner}. Bitcoin, as well as other less popular cryptocurrencies, require their miners to employ devices with high computational capabilities and to be interconnected via the Internet. These requirements play a vital role in the quality and the guarantees of the protocols as well as in the miners' revenue.  All the transactions are stored in a public ledger named \textbf{blockchain} in sets of blocks that are created by the miners\cite{blockchain}. Bitcoin requires from miners to solve cryptographic puzzles, which can only be solved by brute force SHA-256 hashing, in order to generate new blocks for the blockchain \cite{juels1999client}\cite{Rivest:1996:TPT:888615}. Each block has size of 1 MB and two consecutive blocks are created with 10 minutes time difference, on average. 
Miners earn bitcoins whenever they successfully mine a new block and put it in the blockchain. The transactions that are included in a mined block are selected by the miner who successfully mined it. The probability of a miner to mine a block is proportional to the portion of the computational resources of the Bitcoin network he controls. The probability of a transaction to be included in a mined block is proportional to the transaction fees the miner will earn if the block is mined. This gives lower priority to small transactions.

Cryptocurrencies are inferior to conventional currencies, because users cannot exchange money without an Internet connection. 
Mobile wallets are mobile applications that allow anyone who owns credits from a cryptocurrency to create transactions in a similar way as the mobile applications that are offered by banks. 
The role of mobile devices in such scenarios is limited to the submission of the transaction to the authority that maintains the currency, which can be either a decentralised network (Bitcoin) or a mobile phone-based money transfer service (M-Pesa~\cite{mpesa}). 
Mobile devices are practically unable to partake as peers in any decentralised cryptocurrency, because of \textbf{(i)} their lower processing capabilities compared to conventional hardware specialised for mining~\cite{Taylor:2013:BAB:2555729.2555745} and \textbf{(ii)} their unstable connectivity to the Internet compared to ordinary wire-line access protocols. 
For example, consider a university campus which might be spread across an area of some $km^{2}$ with several thousands of users with smartphones. These smartphones can be used to deploy Bitcoin-like currency. However, these devices cannot compete with Bitcoin miners in the block creation process and this gives no incentives to their owners to employ them. Despite that, with widespread usage of smartphones, which are equipped with technologies such as  WiFi-direct, NFC and so on, such users can be easily interconnected.


The problem that we address in this paper is whether we can develop a cryptocurrency, called LocalCoin, that requires neither an Internet connection nor devices with high computational capabilities and is based on the connectivity between users that opportunistically exchange messages. We imagine LocalCoin to complement global-scale cryptocurrencies by handling small transactions without the huge mining expense of Bitcoin-like cryptocurrencies. Localcoin wallets can get charged from Bitcoin wallets through ATM-machine like nodes in the LocalCoin network. 

\subsection{Our Contributions}
We propose LocalCoin, a scheme that replaces the {\it computational} hardness that is at the root of Bitcoin's security with the {\it social} hardness of ensuring that all witnesses to a transaction are colluders (users assisting the malicious user to double spend). Where computational hardness provides a {\em weakest-link} security guarantee - it suffices to break the scheme once - the social hardness provides a {\em strongest-link} guarantee: if just one witness to the transaction is not cooperating, the scheme cannot be broken. This makes it possible to apply the same idea in mobile environments without sufficient computation power or internet connectivity, while taking advantage of its distributed nature~\cite{SytaTVWF15}. 

\textbf{(1)} We are dealing with the proof of ownership issue by proposing a distributed block chain and requiring users to at least store the blocks containing their transactions.

\textbf{(2)} Regarding double spending attacks, we consider the location of each user who verifies the creation of a new block.  
We show that if the network of the users of LocalCoin is dense enough, the probability of a double spending attempt to be successful is upper bounded by the inverse of the square of the number of users (Theorem \ref{thm:pass_attack}). We also prove that the probability of a double spending attempt to be successful by a malicious user who can hire colluders to assist him in the attack, is also very low (Theorem \ref{thm:virtual_attack}). 

\textbf{(3)} We propose an incentive scheme based on transaction and block fees that are adjusted to the ad-hoc networks in order to encourage message exchange.

In addition to the theoretical analysis, we validate our claims regarding the spread of transactions, the transaction rates, and the double spending by extensive experiments on \textit{(i)} static graphs using tools from Random Geometric Graph theory, \textit{(ii)} city scale simulations with mobile users with the help of the ONE simulator~\cite{one} as well as \textit{(iii)} real data from Infocom'05, Infocom'06 and Humanet datasets \cite{cambridge-haggle-2006-01-31,tecnalia-humanet-2012-06-12}.   

\subsection{Applicability of LocalCoin}
We envision LocalCoin as a location-based cryptocurrency that enables small payments\footnote{The technology needed to support the development of LocalCoin is already mature since mobile devices have enough resources and WiFi-direct is supported by every android version since October 2013.}. Although the provided guarantees against double spending are probabilistic and leave a small chance for a double spending attack to be successful, the cost of manipulating the protocol by having a set of colluders in the proper locations outweighs the gains when the transactions are small. As we present in Section \ref{sec:doublespending}, an attacker needs to control many users, who need to be incentivised to attack the protocol and loose their potential earnings from the incentives provided by LocalCoin, in order to split the network in two parts and perform a double spending attack. In order for the chances for a double spending attack to be negligible, the number of the participants has to be high in order to guarantee high connectivity. Given that the transaction verification speed in Bitcoin and other cryptocurrencies depends on the transaction fees that will be collected by the miners~\cite{speedvsfees}, small and local transactions are too expensive and slow to be handled by Bitcoin. LocalCoin can work in parallel with Bitcoin and send transactions to the Bitcoin network that contain many senders and receivers and are practically merged transactions. Technically, a LocalCoin block can be a transaction in the Bitcoin network. 

Apart from conventional money transactions, LocalCoin can also be applied to mobile computing/networking applications such as computation offloading or downloading/streaming services. Device-to-device (D2D) ecosystems have attracted the research interest and various \textit{serverless} architectures and frameworks have been proposed. However, most of them either do not consider incentives for the mobile users that contribute their resources or imply the existence of a centralised server that keeps track of the reliability and the helpfulness of each user. LocalCoin can fill this gap and complement any distributed credit-based incentive scheme for D2D ecosystems.

\section{Related Work}

After Nakamoto's original paper \cite{nakamoto2012bitcoin}, many research groups worked on various perspectives of the Bitcoin protocol. Tschorsch and Scheuermann in their tutorial present the existing contributions and results triggered by the proposal of Bitcoin \cite{7423672}. Garay \textit{et al.} discussed applications, such as the Byzantine agreement, that can be built on top of the Bitcoin core network\cite{Garay2015}. Darkcoin, Zerocoin and CoinShuffle, motivated by the fact that a few transaction deanonymization attacks have been reported, focus on the security and privacy aspects of Bitcoin and propose extensions to fully anonymize transactions \cite{duffield2014darkcoin} \cite{zerocoin} \cite{Ruffing2014}. Also, CoinJoin employs a  multi-signature scheme to enhance the transactions' privacy \cite{maxwell2013coinjoin}. CommitCoin shows a commitment scheme that harnesses the existing computational power of the Bitcoin network \cite{comcoin}. Miller \emph{et al.} present a formal model of anonymous and synchronous processes that communicate using one-way public broadcasts and prove that the Bitcoin protocol achieves consensus in this model in almost any case \cite{miller2014anonymous}. Also Bruce J.D. proposed a cryptocurrency that employs a mini-blockchain with finite number of blocks in order to reduce the storage requirements and improve efficiency~\cite{bruce2013purely}.

Authors of \cite{Karame12twobitcoins} analyse the security of using Bitcoin for fast payments, where the time between the exchange of currency and goods is short. Furthermore, \cite{fast} investigates the restrictions on the transaction processing rate in Bitcoin as a function of both the bandwidth available to users and the network delay, both of which lower the efficiency of Bitcoin's transaction processing. The security analysis done by Bitcoin's creator assumes that block propagation delays are negligible compared to the time between the creation of two consecutive blocks. This assumption fails when the protocol is required to process transactions at high rates. Eyal \textit{et al.} proposed Bitcoin-NG, the `next generation' of Bitcoin the design of which is based on scalability \cite{194906}. In more detail, the latency is limited only by the propagation delay of the network and the bandwidth of the capabilities of the miners. 

Moreover, \cite{DBLP:journals/corr/EyalS13,BMPACGTA} argue that the Bitcoin protocol is not incentive-compatible and after presenting a game theoretic analysis they also present an attack in which colluding miners obtain a revenue larger than their fair share. Also, \cite{TaMDMiBC} introduces a new defence against this 51$\%$ attack via (i) presenting a block header, (ii) introducing some extra bytes, and (iii) utilising the time-stamp more effectively in the hash generation. According to \cite{DBLP:journals/corr/DeckerSW14}, Bitcoin only provides eventual consistency. They propose PeerCensus, a new system, built on the Bitcoin block chain, which enables strong consistency and acts as a certification authority, manages peer identities in a peer-to-peer network, and ultimately enhances Bitcoin and similar systems with strong consistency. 

\section{Proposed Approach}\label{sec:featuresOfCryptocurrencies}
As discussed earlier,  cryptocurrencies have to address three main challenges. We, first, explain how Bitcoin addresses these challenges and then how LocalCoin encounters them.

\subsection{Bitcoin}

\textbf{Proof of ownership:} Bitcoin's main achievement is its ability to reach a consensus about a valid transaction history in a totally decentralised fashion. Bitcoin deals with the proof of ownership problem by using the concept of \textit{block chain} based on a Merkle tree data structure. The block chain consists of a sequence of blocks connected in a hash chain, where every block imprints a set of transactions that have been collected from the network. Every miner is aware of the creation of a new block and consequently is able to validate the proof of ownership of a claimed Bitcoins. 
Users can employ their bitcoins by using a set of verified transactions. In order for one transaction to be counted as verified it has to belong to a block which is at least six blocks away from the current mined block in the block chain. 

\textbf{Double spending avoidance:} Bitcoin overcomes double spending by using a proof-of-work mechanism that imposes a delay on the verification of the transaction. In order to overcome this mechanism, one has to solve a hard problem with input that takes approximately 10 minutes for a brute force algorithm to solve. 
There are three main ways to attempt double spending in Bitcoin protocol: (\textit{i}) race attack, (\textit{ii}) Finney attack and a (\textit{iii}) 51$\%$ attack. Waiting for some new blocks to be created based on the current one can easily prevent the first two attacks; One block in the case of a race attack and six in the case of a Finney attack. However a 51$\%$ attack can collapse the whole Bitcoin network but this is extremely costly.  
Also, Eyal and Sirer prooved that proof-of-work blockchains are vulnerable to selfish mining by attackers that control more than 1/4 of the network's mining power \cite{DBLP:journals/corr/EyalS13}. 

\textbf{Incentives:} Each miner gets a reward of 25 bitcoins for mining a block \cite{bitcoinI}. However, this reward halves every 4 years. Another concern is, as more and more users join as miners, the probability of mining a successful block reduces. To partially overcome this issue, miners create mining pools and they share the earnings whenever one of them solves a cryptographic puzzle \cite{eyal2015minersDilemma}. 

\subsection{LocalCoin}\label{sec:contributions}

In this work, we propose a new Bitcoin-like cryprocurency protocol, namely LocalCoin, for mobile ad-hoc networks in urban areas with high device density. 

\textbf{Proof of ownership:} LocalCoin uses a \emph{lightweight} storage architecture by extending the concept of  blockchain in a distributed fashion, where each user can store as many blocks as she wants. The proposed distributed blockchain has a redundancy factor between the users. LocalCoin, similarly to Bitcoin, stores transactions into blocks. 
All the transactions in the same block are collectively verified. \textit{The size of each block} is denoted by $BS$. In order for one block to be created \textit{a minimum number of users to verify each transaction}, denoted by $\mathit{mVu}$, is needed (i.e., at least $BS \cdot \mathit{mVu}$ users are informed about each transaction on one block). 
The relationship of these variables with the total amount of users affects the time needed to verify one block and prove the ownership of all the users that own these transactions.

\textbf{Double spending avoidance:} LocalCoin nullifies Bitcoin's computation overhead via incorporating a novel protocol, which is designed for the ad-hoc environment. Bitcoin's proof of work is based on the fact that cheating is improbable because a malicious user has to solve hard problems at a faster rate than the total remaining users. In LocalCoin, cheating is made very difficult because a malicious user has to misinform the majority of a set of trusted users. Every user in the LocalCoin protocol selects the users she trusts. LocalCoin avoids double spending in two ways. \textbf{(i)} The receiver of one transaction will accept the transaction if and only if she receives the transaction signed by at least \textit{a minimum number of trusted users} of her trusted network, denoted by $\mathit{mTr}$. This constraint imposes a useful delay that spreads the transaction message to more users and increases the probability of one trusted user to detect the same input to another transaction. 
It is worth mentioning that any initiated transaction is signed by the sender and we assume that it is impossible for a malicious user to fake a transaction by pretending to be another user. \textbf{(ii)} During the block creation process, every participant checks for double spending attempts. To avoid fake block creation attempts by a set of collaborative malicious users, LocalCoin enforces the \textit{average distance between the users that will verify the creation of a new block} to be more than $\mathit{aVd}$. This last constraint allows the block creation messages to be scattered to as many users as possible. 

\textbf{Incentives:} We extend the transaction fee schema in order to motivate mobile users to participate. We propose \textit{transaction fees} to motivate users to forward messages and \textit{block fees} to motivate them to store as many blocks from the distributed block chain as possible. Transaction fees are important because mobile users are competing for them and they broadcast any received transaction. Every transaction includes an amount of localcoins that are collected by the mobile user who will first inform the receiver of the transaction about the transaction. The probability of a mobile user to earn the transaction fees does not depend on the technical characteristics of their mobile device since D2D communication protocols perform similarly in different devices. Block fees are important because users store the created blocks in order to be able to verify the creation of new ones. Whenever a block is created, the mobile users that verified each transaction because they where aware of it share the localcoins that were included in these transactions as block fees. 
 
LocalCoin is a lightweight protocol because: \textbf{(i)} any user has to only forward messages, \textbf{(ii)} a small subset of the total users are checking for the validity of a transaction message and \textbf{(iii)} users have to store the blocks that include their own transactions. The feasibility of LocalCoin is proved using concepts from random geometric graph (RGC) theory and its performance is depicted with the help of static graphs (Section \ref{sec:matlabsims}), real trances from mobile users (Section\ref{sec:javasims}) and simulations with mobile users in a city scale (Section \ref{sec:onesims}). In order to explain LocalCoin, we assume that it is deployed as a service in an area\footnote{Terms "service" and "protocol" appear in this paper and the former is an implementation of the later.}.

\begin{figure}[H]
    \centering
\begin{subfigure}{\columnwidth}
\includegraphics[width=\columnwidth]{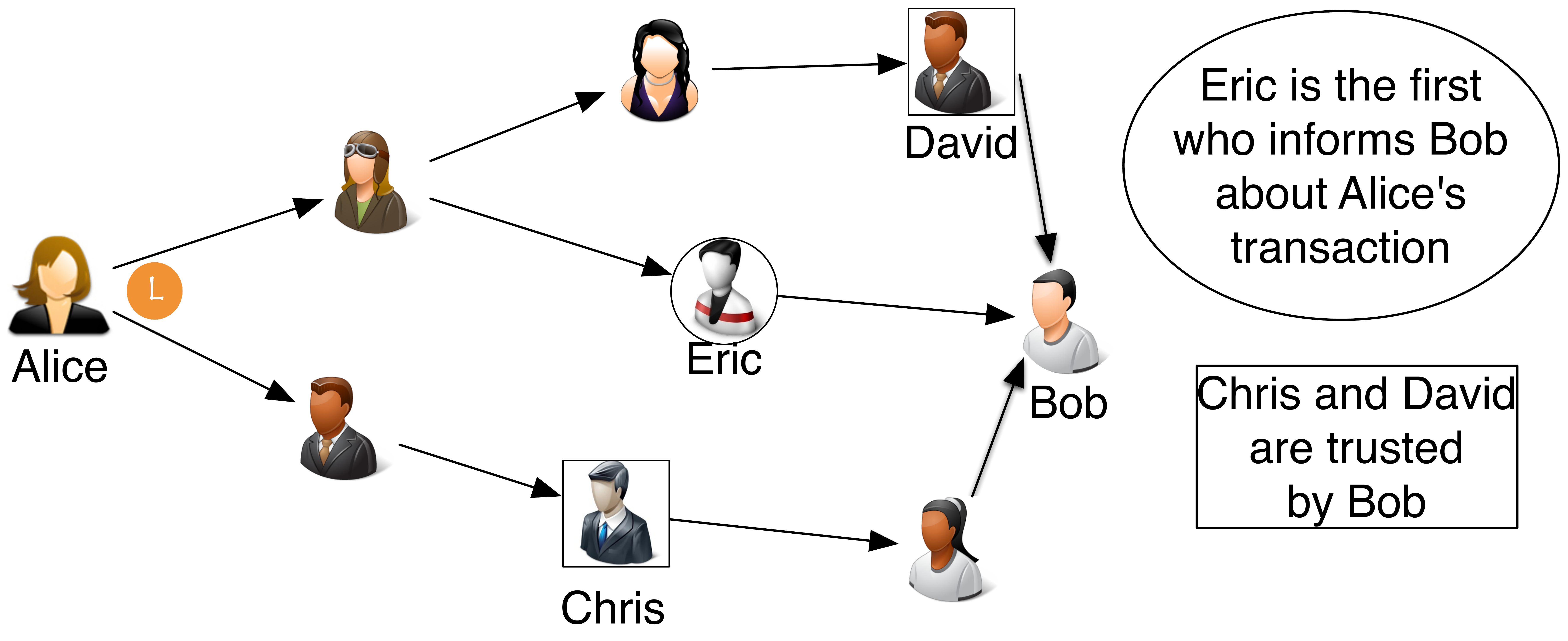}
    \caption{Alice transfers localcoins to Bob ($\mathit{mTr} = 2$). Eric gets the transaction fees.}   
    \label{fig:f1}
\end{subfigure}
\begin{subfigure}{\columnwidth}
\includegraphics[width=\columnwidth]{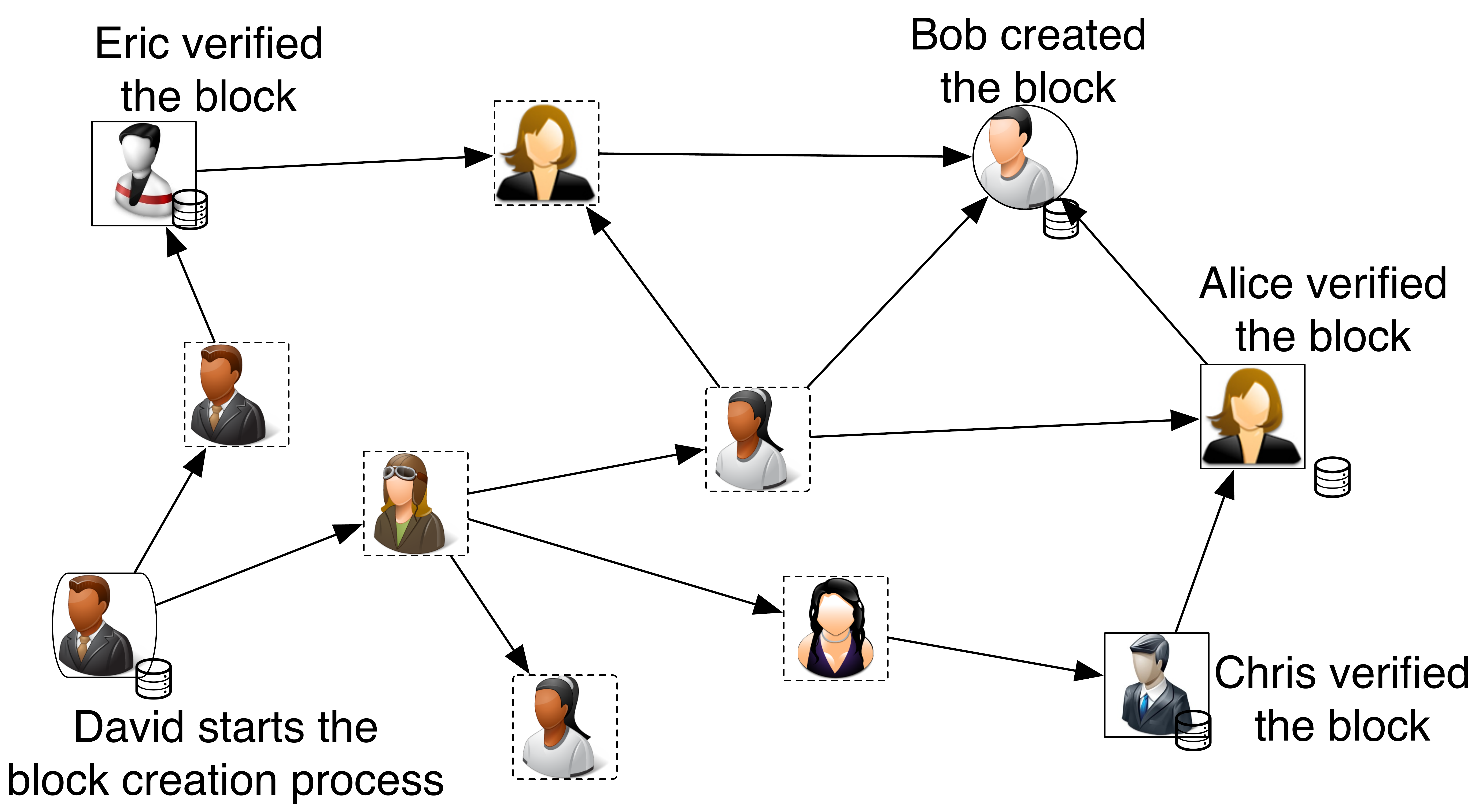}
    \caption{David builds a new block, Chris, Eric and Alice verify it and Bob creates it. ($\mathit{mVu}$ = 5, Average distance between David, Chris, Eric, Alice and Bob $> \mathit{aVd}$).}   
    \label{fig:f2}\end{subfigure}
    \caption{Transaction and Block creation examples.\vspace{-0.3cm}}
    \label{fig:example}
\end{figure}

\textbf{Examples:} Figures \ref{fig:f1} and \ref{fig:f2} depict the broadcasting of a transaction and a block creation. Alice broadcasts $t_{Alice\to Bob}$ to her neighbors who forward $t_{i \to j}$ because they hope to get the transaction fees. Their neighbors also forward the transaction for the same reason and then Eric is the first who informs Bob. If we assume for this example that $\mathit{mTr} = 2$, after the reception of $t_{Alice \to Bob}$ with the signatures of Chris and David, Bob will broadcast his ack message and he will announce Eric as the receiver of the transaction fees. This pair of messages will be stored by at least Alice, Bob and Eric and potentially more users that participated in the forwarding and will be used in the block creation process. After collecting $BS$ transaction pairs, David broadcasts a block creation message, Alice, Chris and Eric verify the block and Bob creates it since the average distance between them is more than $aVd$ and $\mathit{mVu} = 5$. 

\section{The settings}\label{sec:probDesc}

We assume a set of mobile users $\mathcal{U}$ who are registered to LocalCoin service. Every user is self-interested and can be malicious. Each user $i \in \mathcal{U}$ can utilise the service if she is inside the geographical area, $d_{i} \in \mathcal{D}$ and can change $d_{i}$ only by moving to another location and not by manipulating it. We discuss further this assumption in Section \ref{sec:assumptions}. Two users can communicate only if their distance is within a threshold, so if a malicious user tries to manipulate his location, he will be detected. 
Any user $i \in \mathcal{U}$ is able to exchange localcoins with another user $j \in \mathcal{U}$ by creating one transaction $t_{i \to j}$. User $i$ needs to broadcast the transaction message that determines its characteristics. Any user $j$ has a set of trusted users $\mathit{TN}_j$ and this selection is based on social interaction between users as well as on other device to device interactions \cite{socialtrust}. A realistic requirement of our protocol can be to force users to select their trusted peers by pairing via NFC.
The selection of $\mathit{TN}_j$ depends on $j$ and they are responsible for guaranteeing that any received transaction with $j$ as a destination should be examined before being broadcasted. The forwarding procedure is explained in detail in Section \ref{sec:transmessages}. User $i$ owns some localcoins and in order to prove this ownership she remembers all the transactions in which she was the receiver. The transactions are stored in blocks and each block contains more than one transaction. Each block is based on a previous block by creating a blockchain. We call this chain \textit{distributed blockchain} because each block is duplicated to more than one but not to every user. 

\begin{table}[t]
\normalsize
    \begin{center}
    \begin{tabular}{ l  p{1.3 cm} p{1.3 cm} p{1.3 cm} p{0.3 cm} p{0.3 cm}}
    \toprule
    \textbf{Input:}  	& $h(t_{* \to i}(1))$  	& $h(t_{* \to i}(2))$  	& $h(t_{* \to i}(3))$			& $\ldots$ &    \\ 
    \cmidrule(r){1-6}
    \textbf{Output}	& $o_{j}$			& $o_{i}$			& $\mathit{\mathit{trf}_{ij}}$	& $\mathit{\mathit{bf}_{ij}}$ & $b_{i}$ \\
    \bottomrule 
      \end{tabular}
      \vspace{-0.2cm}
    \caption{Transaction Template. \vspace{-0.5cm}}
	\label{tab:trans}
	\end{center}
\end{table}

The mobile ad-hoc network nature of our service allows any other user $k$ in the area to detect the transaction message, collect the information and contribute to this transaction. Any collected third party transaction can be used in the future by user $k$ to earn money in terms of transaction fees and block fees. Transaction fees motivate mobile users to forward any received transaction message while block fees motivate mobile users to store collected transaction messages. We assume that any user $k$ is self-interested and participates in the system in order to earn localcoins and use them later. Users that are not willing to sacrifice part of their resources in order to participate and earn localcoins are not considered since they only co-exist with the ones that participate. The more the active mobile users the better for the protocol and for that reason we design two types of incentives and also allow mobile users to stop participating if they want to save their resources and rejoin again later. We also consider malicious users that want to attack the protocol and double spend their localcoins. The robustness of Localcoin against the considered attacks is presented in Section~\ref{sec:doublespending}.
Each transaction $t_{i\to j}$ is described by a set of inputs and outputs as presented in Table \ref{tab:trans}.  $h(t_{* \to i})(\cdot)$ is the hash of a block that contains a transaction from anyone to user $i$. The outputs are: the transferred amount to user $j$, $o_j$, the transaction fees $\mathit{trf}_{i \to j}$, the block fees $\mathit{bf}_{i \to j}$, any possible change $o_{i}$ and the amount of money user $i$ owns, $b_{i}$. Bitcoin employs SHA-256 hashing algorithm, by adapting it in LocalCoin the size of each transaction will be $32 \cdot \#input\_transactions + 160$ bytes. A transaction with less than 10 transactions as input requires less than a kilobyte of storage. 
Transactions are verified in blocks via a mechanism that is presented in Section \ref{sec:blkcrmessages}. Observe that a transaction $t_{i \to j}$ additionally: (i) returns, if any, change to i, (ii) transfers transaction fees to appropriate users, and (iii) pays block fees if needed. From now on, whenever we are using a past transaction as an input to a new one, where user $i$ transfers some money, we assume that $o_i$ can be of any of the four previous types. 
Each user $i$ keeps a transaction database $\mathcal{T}_{i}$, which contains a subset of the distributed block chain and a set of pending/unverified transactions. This database contains all the blocks user $i$ needs to verify her own localcoins $\mathcal{B}_i \subseteq \mathcal{T}_i$ as well as other blocks in which she was present. At any time, there are two types of transactions in the network, the verified ones that can be used as an input to a new transaction and are stored in the blockchain and the unverified ones. Unverified transactions are verified in bunches by a distributed consensus protocol and added to the distributed blockchain.

\section{Protocol}\label{sec:block_creation}
We present the basic functionalities of the LocalCoin protocol, which are categorised into three main categories; transaction messages (Section \ref{sec:transmessages}), block creation messages (Section \ref{sec:blkcrmessages}), and block management messages (Section \ref{sec:blkmanmessages}).

\subsection{Transaction Messages}\label{sec:transmessages}

\begin{algorithm}[t]
\caption{Pseudocode of the transaction processes}
\label{alg:transcode}
    \begin{algorithmic}[1]
    \Procedure{Send$\_ t_{i \to j}$}{}
        \LState $t_{i \to j}$=create$\_$transaction(), t=sign($t_{i \to j}$)
        \LState broadcast(t,$my\_id$)
    \EndProcedure
    \end{algorithmic}

    \begin{algorithmic}[1]
    \Procedure{receive($t_{i\to j}$,$k$)}{}
    \If{$t_{i\to j}$ is a send message and is new}
        \If{$my\_id$ == j}
            \LState process($t_{i\to j}$,$k$)
        \Else
            \If{$j \in \mathit{TN}_{my\_id}$}
                \LState $aware\_of\_blocks \gets check(t_{i\to j})$
                \If{$aware\_of\_blocks == true$}
                    \LState t=sign($t_{i \to j}$)
                \EndIf
            \EndIf
            \If{$k \in \mathit{TN}_{my\_id} \text{ and } t_{i\to j} \text{ signed by }k $}
                \LState $update\_my\_blockchain()$, t=sign($t_{i \to j}$)
            \EndIf
            \LState broadcast(t,$my\_id$), Update$\_$pending$\_$List()
        \EndIf
    \Else
        \LState Match$\_$pending$\_$List(), $pending \gets pending+1$
    \EndIf
    \EndProcedure
    \end{algorithmic}

    \begin{algorithmic}[1]
    \Procedure{process($t_{i\to j}$,$k$)}{}
        \If{$t_{i\to j}$ is new}
            \LState $first\_notifier(t_{i\to j}) \gets k$, $trusted(t_{i\to j}) \gets 0$
        \EndIf
        \If{$k \in \mathit{TN}_{my\_id}$}
            \LState $trusted(t_{i\to j}) \gets trusted(t_{i\to j})+1$
       \EndIf
        \If{$trusted(t_{i\to j})> \mathit{mTr}$}
            \LState ack($i,j,t_{i \to j},first\_notifier(t_{i \to j})$)
        \EndIf
    \EndProcedure
    \end{algorithmic}
\end{algorithm}

\textit{send(i,j,t$_{i \to j}$)}: User $i$ broadcasts a transaction, as described in Table \ref{tab:trans}, in order to transfer a number of localcoins to user $j$. The \textit{send} command will broadcast $t_{i \to j}$ to all the nearby users (\textit{neighbors}). Any user $l$ operates based on the functionality of the \textit{receive(t$_{i \to j}$)} procedure as described in Algorithm \ref{alg:transcode}. Whenever she receives a new transaction, she checks the sender and receiver and if she is not familiar with either of them she forwards the message hoping to collect the transaction fees. If she belongs to the trusted users of the receiver of the transaction, she examines the input transactions and signs the message if she is able to validate all of them. If she receives the message by a trusted user, she updates her transaction database according to the signed message. If she is the receiver of the message, she processes it as explained in procedure \textit{process(t$_{i \to j}$,k)}. If she received the same transaction before, she ignores the message unless it is now signed by another trusted user of $j$. After receiving the transaction, $j$, has to wait for at least $\mathit{mTr}$ of her $\mathit{TN}_j$ trusted users to sign and forward the transaction to her.
The first user who forwards this message to $j$, regardless of being in her trusted users, will receive the amount of $\mathit{trf}_{ij}$ if the transaction is going to be accepted by $j$ and verified by the network. By accepting the transaction only if a subset of the trusted network signs and forwards the message to the receiver, the protocol addresses sybil attacks. User $i$ will not stop broadcasting the transaction to any users she meets until she will receive the \textit{ack} message from user $j$. 

\textit{ack(i,j,t$_{i \to j}$)}: If $j$ receives the message from $\mathit{mTr}$  users of her trusted network $\mathit{TN}_j$, then she broadcasts an acknowledgement message. This message also determines the address of the user that will receive the transaction fees. This message is also forwarded in the similar way to that of the \textit{send} command. There is no need for a third round because user $j$ can only assign $\mathit{trf}_{ij}$ to someone else. Everyone who receives the acknowledgement updates the knowledge about which accounts are participating in the transaction. 

\begin{algorithm}[t]
\caption{Pseudocode of block creation processes}
\label{alg:blockcode}
    \begin{algorithmic}[1]
        \Procedure{BuildBlock}{\textbf{BLK}($t_{i \to j},t_{i^{'} \to j^{'}},\ldots$)}
            \LState $block \gets t_{i \to j},t_{i^{'} \to j^{'}},\ldots$, $locations \gets 0$
            \LState $locations(0) = my\_GPS$
            \LState broadcast($block,locations$);
        \EndProcedure
    \end{algorithmic}

    \begin{algorithmic}[1]
        \Procedure{Receive}{$block,locations$}
            \LState $[is\_valid,signed] \gets Verify(block)$
            \If{$is\_valid$}
                \LState $avdist \gets Average\_Distance(locations)$
                \If{$avdist > \mathit{aVd}$}
                    \LState $broadcast(created\_block)$
                \EndIf
            \Else
                \LState broadcast("double spending attempt detected")
            \EndIf
        \EndProcedure 
    \end{algorithmic}
\end{algorithm}

\subsection{Block Creation Messages}\label{sec:blkcrmessages}

\textit{build}(\textbf{BLK}($t_{i \to j},t_{i^{'} \to j^{'}},\ldots$)): Whenever a user $l$ collects $BS$ transactions (both \textit{send} and \textit{ack}), that are not yet verified, she 
tries to build a new block. For that she needs to agree with $\mathit{mVu}-1$ other users about the validity of the $BS$ transactions in order to reach to a consensus~\cite{ConsensusProblem}. Her signed message also contains her location, $d_l$ and a current value of average distance vector $\textbf{d}$. The distance vector has $BS$ entries and each entry has the average distance between the users who verified the transaction.

\textit{verify}(\textbf{BLK}$^{'}$($t_{i \to j},t_{{i}^{'} \to {j}^{'}},\ldots$)):  The first $\mathit{mVu}$ users who verify all the transactions in the \textit{create} message and have average distance between each other bigger that $\mathit{aVd}$ will share the block fees. Every user $k$ who receives a \textit{verify} message checks her database for unverified transactions and if she has any of the included in the message she signs them and forwards the message. Before forwarding the message, user $k$ updates the distance entries, which she has signed. If she detects a double spending attempt she deletes her entry if it has a later time-stamp or she signs her entry and adds it into the message if it has an earlier time-stamp. In case of double spending detection, user $k$ sets the entry for the corresponding transaction to $0$ and attaches and signs her detected pair with a newer timestamp. Whenever a user receives a \textit{verify} message with the location of the user not being in her coverage radius, \textit{(i)} she verifies any transaction she can verify, \textit{(ii)} she notes that the location of the receiver is not in her coverage radius and she marks the location entry as false and then \textit{(iii)} she broadcasts the message.

\textit{create}(\textbf{BLK}$^{''}$($t_{i \to j},t_{{i}^{''} \to {j}^{''}},\ldots$)): Users who receive a message with transactions that are verified $\mathit{mVu}$ times and have average distance bigger than $\mathit{aVd}$ apart from forwarding the message they also broadcast a \textit{create} message that defines the users who will share the block fees. Before broadcasting the \textit{create} message, they examine if there is any entry of the $\mathit{mVu}$ that has been marked by another user as false and in such case, these entries are not considered in the block creation.

\subsection{Block Management Messages}\label{sec:blkmanmessages}

\textit{delete(i,$t_{* \to i}$)}: We propose a garbage collection functionality that deletes every transaction that is not useful. The \textit{delete} command is triggered after the \textit{create} command in order to delete all the input transactions to the freshly verified ones since they can not be used any more. After deleting all the transactions of one block, the whole block is deleted. The motivation behind this process is to keep the size of the distributed block chain as storage efficient as possible because the mobile devices are not able to dedicate significant amount of storage for that\footnote{Bitcoin's blockchain is increasing with a rate of more than 200 MB per day \cite{blockchainSize}.}. 

\textit{sync(t)}: Any user can call \textit{sync} function by giving only the time-stamp of her last update. By doing so, any nearby trusted user will send the newly verified transactions as well as the hash of the ones that have been deleted.

\begin{figure}[t]
    \centering
    \includegraphics[width=\columnwidth]{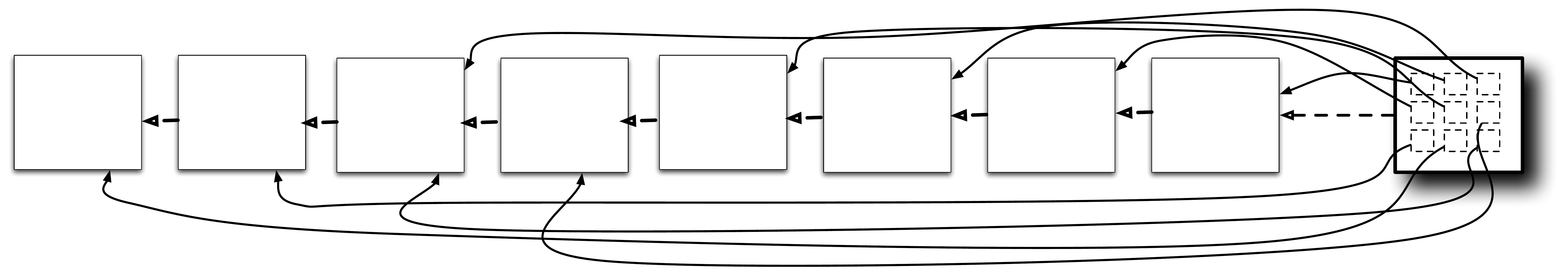}
    \caption{A zoom view of one block with the pointers of each transaction to its parent block(s). For the sake of presentation we show that each transaction has only one parent.\vspace{-0.3cm}}
    \label{fig:chainzoom}
\end{figure}

\subsection{The Blockchain Evolution}
If we have $\Lambda$ transaction pairs per time unit, then we will have, in the long term, $\Lambda/BS$ blocks per time unit. If one transaction $t_{i \to j}$ uses on average $L_{i \to j}$ transactions as an input then one block deletes $\sum_{k = 1}^{BS}L_{i \to j}^{k}$ links to past blocks. In order to become one block orphan, all its transactions need to be unpointed. Each transaction is pointed by 4 links, so this block will be deleted approximately when $4*BS$ transactions that point to that block are deleted. However, we cannot predict after how many block creations this will happen. Whenever a new block is created, the garbage collection process updates the past blocks on which the inputted transactions where placed. For any used transaction, in the creation of the new block, we delete the pointers to the parents of the used transactions. Figure \ref{fig:chainzoom} is a pictorial view of where one transaction of the nine in the rightmost block is used as an input to a new transaction. All the links to the parent blocks of this transaction will then be deleted. If a block has no child pointers pointing at it, it is deleted and the block that is after it then points to the one before it.


\subsection{Parameters of LocalCoin}
The performance of LocalCoin depends on multiple parameters. User availability and position determine the connectivity between the users and the ability to verify new transactions but these parameters are not regulatable by LocalCoin. On the other hand, the number of the trusted users needed for one transaction to be accepted ($\mathit{mTr}$), the amount of the transactions in each block ($BS$), the number of the users need to verify the creation of one block ($\mathit{mVu}$) and the average distance between the users who verify the block ($\mathit{aVd}$), affects the performance of LocalCoin and characterises the trade-off between the time needed to verify a new transaction, the security against double spending and the increase in the stored data. However, since LocalCoin can be categorised as a location based service, all four parameters can be adjusted based on the required transaction speed and the maximum risk of double spending as well as the rate of the creation of a new block.

\section{Analysis}\label{sec:analysis}

In this section we analyse and validate the performance of LocalCoin. Section \ref{sec:reachable} introduces the connectivity conditions under which the protocol is applicable and Section \ref{sec:doublespending} presents the circumstances where a malicious user is able to successfully double spend some localcoins. Given that users' mobility increases the capacity of multihop wireless networks \cite{Grossglauser:2002:MIC:581862.581866}, we examine the worst-case performance of LocalCoin by considering non moving users.

\subsection{Reachability}\label{sec:reachable}
Each user $i$ is located at $d_{i} \in \mathcal{D}$ and any other user $j$ located in the coverage area of $i$, $D_i$ (i.e $j \in D_i$) is able to receive any message $i$ broadcasts. Practically, the coverage area of each user has a radius comparatively close to the coverage area of WiFi direct. We assume that every user has the same normalised coverage radius and we denote it as $r_{cov}=\frac{Wifi\_direct\_coverage}{Area\_of\_the\_supported\_service}$. Given the location of each user and the normalised coverage radius we produce a $2$-dimensional random geometric graph (RGC) $G_{\mathcal{D}}(\mathcal{U},r_{cov})$ . The number of connected components of $G_{\mathcal{D}}(\mathcal{U},r_{cov})$ depends on $|\mathcal{U}|$ and $r_{cov}$ and has a subcritical and a supercritical phase. In the subcritical phase the number of connected components is large while in the supercritical phase it converges to one. A well known result for $d$-dimensional random geometric graphs is the following \cite{gupta2000capacity,penrose2003random}: 

\begin{lemma}
For $|\mathcal{U}|r_{cov}^{d} \geq 2\log |\mathcal{U}|$ the $G_{\mathcal{D}}^{d}(|\mathcal{U}|,r_{cov})$ is \\
	\textbf{(1)} connected with probability at least $1-\frac{1}{|\mathcal{U}|^2}$,  \\
    \textbf{(2)} $r$-regular and \\
    \textbf{(3)} the degree of every user, with high probability, is $\frac{\pi^{d/2}}{\Gamma(1+d/2)}nr^{d}(1+o(1))$.
\end{lemma}
Where $\Gamma(\cdot)$ is the Gamma function and given that we consider a 2-dimensional graph ($d=2$), $\Gamma(2) = 1$. We can rewrite the lemma as: If $\frac{|\mathcal{U}|}{\log |\mathcal{U}|} \geq \frac{2}{r_{cov}^2}$, $G(|\mathcal{U}|,r_{cov})$ is regular with degree $d_{\mathcal{D}}=\pi |\mathcal{U}| r_{cov}^{2}(1+o(1))$. 
Given a $\beta$-expander $d_{\mathcal{D}}$-regular graph, for every set $S \subset \mathcal{U} , |S| \leq |\mathcal{U}|/2$, holds $out(S) \geq \beta |S|$. Where:
$out(S) = \left| \{ \{u,v\} | \{u,v\} \in G_{\mathcal{D}}(|\mathcal{U}|,r_{cov}) , u \in S, v \notin S  \} \right|$

Authors of  \cite{kong2007critical,1498550} state that if the nodes of the random geometric graph are produced by Poisson point process in the 2 dimensions, its density should be in the spectrum of $[0.696, 3.372]$. Simulation results converge to $1.44$. If for example the subscribed users are 1000 and the coverage radius of wifi-direct is 200 meters the users will be able to form a connected graph with probability 0.999999 if the area of the supported service is less than $\pi \sqrt{\frac{2}{3}10^{11}} \approx 0.8 km^2$.

Suppose user $i$ wants to give some localcoins to user $j$. User $i$ has $d_{\mathcal{D}}$ neighboors and by the properties of the expander graphs, there are $(1 + d_{\mathcal{U}})(1 + \beta)^l$ users at $l$ hops from $i$. We continue expanding from $i$ until the reachable set of users $V_i$ has more than $|\mathcal{U}|/2$ users. User $j$ may not be among them. However, if we expand from user $j$ in the same way, we eventually obtain a set $V_j$ of more than $|\mathcal{U}|/2$ users reachable from $j$. The sets $V_i$ and $V_j$ both contain more than $|\mathcal{U}|/2$ users so they must overlap. The overlap contains users on a path from $i$ to $j$. In this way, we have shown that:
\begin{theorem}\label{them:pathexists}
For any pair of users $i$ and $j$, in the same connected component, there is a path of length at most $2(l + 1)$ from $i$ to $j$, where $l = log_{(1+\beta)}\frac{\mathcal{U}}{2d_{\mathcal{D}}}$.  The larger the value of $\beta$, the shorter the path between any two users.
\end{theorem}
In LocalCoin protocol, $j$ has to be connected with at least $\mathit{mTr} \in \mathit{TN}_{j}$ in order to accept the transaction. Theorem \ref{them:pathexists} ensures that if user $i$ wants to transfer some localcoins to user $j$ the only requirement is that user $j$ be connected with at least $\mathit{mTr}$ users of her trusted network. The probability of the transaction to be successful depends on two main factors. The most important factor is both $i$ and $j$ must belong in the same component $c$, that is : $p(i \in c)\cdot p(j \in c)=|\% c|\cdot |\% c| = |\% c|^2$. Where $|\% c|$ is the fraction of the users that belong to component $c$. The second factor is to have at least $\mathit{mTr} \subset \mathit{TN}_j$ of $j$'s trusted users in the component. This probability equals:
\begin{align}
\sum_{l=  \mathit{mTr}}^{|\mathit{TN}_j|}{\mathit{TN}_j \choose l}(p\cdot|\% c|)^l(1-p\cdot |\% c|)^{\mathit{TN}_j - l}
\end{align}
Where $p$ is the probability of one user who belongs to the trusted network of $j$ to be able to sign $i$'s message. It is worth mentioning that in the case of moving users, the probability of having a successful transaction is increasing because mobile users can forward the received transactions whenever they make new neighbours. 

\begin{table}[t]
\normalsize
    \begin{center}
    \begin{tabular}{ l  p{6.5 cm} }
     \toprule
     \textbf{Symbol}      &  \textbf{Meaning}    \\
      \cmidrule(r){1-2}
     $d_{i}$	&	Location of user $i$ in area of consideration $\mathcal{D}$. \\
     $r_{cov}$	& 	Normalised coverage area.	\\
     $\mathcal{U}(A)$ & The users that are located in area $A$. \\	
     $\mathcal{M}$ & Set of colluders. The users that help a malicious user $m$ to double spend a localcoin. \\
     $\mathcal{R}$ & The area that is controled by malicious user $m$ who tries to double spend a localcoin. \\
    \bottomrule \vspace{-0.5cm}
    \end{tabular}
    \caption{Notation table with frequently used symbols. \vspace{-0.5cm}}
	\label{tab:notation}
	\end{center}
\end{table}


\subsection{Robustness Against Attacks}\label{sec:doublespending}

Malicious users may try to attack LocalCoin in various ways and in case of a successful attack they may be able to steal localcoins from other users or just harm the protocol. There are two general categories of attacks in LocalCoin, the ones that are based on the proof-of-ownership and the ones that are based on double spending. In this Section we focus on malicious users who want to earn localcoins via double spending since this is the most popular attack on cryptocurrencies. 
Attacks where mobile users input fake transactions and try to send them to other users will not be successful since the trusted users of the recipient will not sign them but such attacks will consume the resources of the mobile users that forwarded them. 
Let us assume that a malicious user $m$ wants to double spend a localcoin. We consider two types of attacks: 

\textbf{Passive:} The attacker initiates two transactions to two different receivers and broadcasts them to two different connected components of the network. If there is only a single connected component, the attempt will be detected and hence there must exist at least two disjoint components for $m$ to be successful in double spending. In Section \ref{sec:coopAttack} we show that probability of such attack decreases quadratically in $|\mathcal{U}|$.

\textbf{Active:} Another possible attack is the one where $m$ is able, with the help of some colluders ($\mathcal{M}$), to control an area $\mathcal{R}$ and split $\mathcal{D}$ into more than one disconnected parts artificially. We discuss such attack in Sections \ref{sec:splitAttack} and \ref{sssec:vc_dn}. 

In both types of attacks we assume that the trusted users who are asked to verify transactions are aware of every block.  

\subsubsection{A Passive double spending attack}\label{sec:coopAttack}
In order for double spending to be successful, $m$ has to employ enough colluders in order to cheat at least $2\mathit{mVu}$ other users. Each recipient of the fake transaction will wait until $\mathit{mVu}$ of her trusted nodes will forward her the fake message. Depending on $BS$, $\mathit{mVu}$ and $\mathit{aVd}$ the probability of double spending is changing. However, the wireless medium does not allow $m$ to only select a number of users. Given that $\mathit{mTr}$ users from both receivers are aware of $m$'s ability to initiate this transaction, we examine how the remaining parameters affect the difficulty of double spending:

\textbf{$BS$:} The lower the number of transactions in one block the faster each block can be created and this allows $m$ to try to double spend the same input and create two new blocks. If the connectivity graph between the users is partitioned, double spending is possible.

\textbf{$\mathit{mVu}$:} The higher the number of users needed to verify one transaction the more $m$ needs to collaborate with him. For that, Lemma 1 can not hold and the connectivity graph has to be in the subcritical phase.

\textbf{$\mathit{aVd}$:} The higher the value of $\mathit{aVd}$ the most difficult it is for $m$ to double spend. Each user has on average $d_{\mathcal{D}}$ neighbours and any two users can communicate if the distance between them is less than $r_{cov}$, then if $\lambda = \mathit{aVd}/r_{cov}$,  $\lambda \cdot d_{\mathcal{D}}$ users will receive the request for fake block creation. 

We can paraphrase Lemma 1 and state that:

\begin{theorem}\label{thm:pass_attack}
\begin{align*}
& \text{If }  |\mathcal{U}| \left(\frac{\mathit{aVd}}{\lambda}\right)^2 \geq 2\log |\mathcal{U}| \text{ and }   \lambda d_{\mathcal{D}} > \frac{|\mathcal{U}|}{2}, \text{double}\\
& \text{spending is possible with probability at most } 1/|\mathcal{U}|^2.
\end{align*}

\end{theorem}

\subsubsection{Virtual-cut attack: An active double spending attack in static graphs}\label{sec:splitAttack}
Let us assume that the users' positions, $\{d_i\}$, are distributed uniformly and are static. As proved in the previous subsection, it is difficult to double spend a localcoin in the induced random graph as with high probability it consists of one major component. However, a malicious user may have detailed knowledge of the graph topology and he may artificially create a \emph{virtual cut} by controlling users that transmit messages selectively to one part of the graph only. If a malicious user is able to double spend a localcoin by such trick, we say he is successful in a \emph{virtual-cut attack}. 

\begin{figure}[t]
	\centering
    \includegraphics[width=0.7\columnwidth]{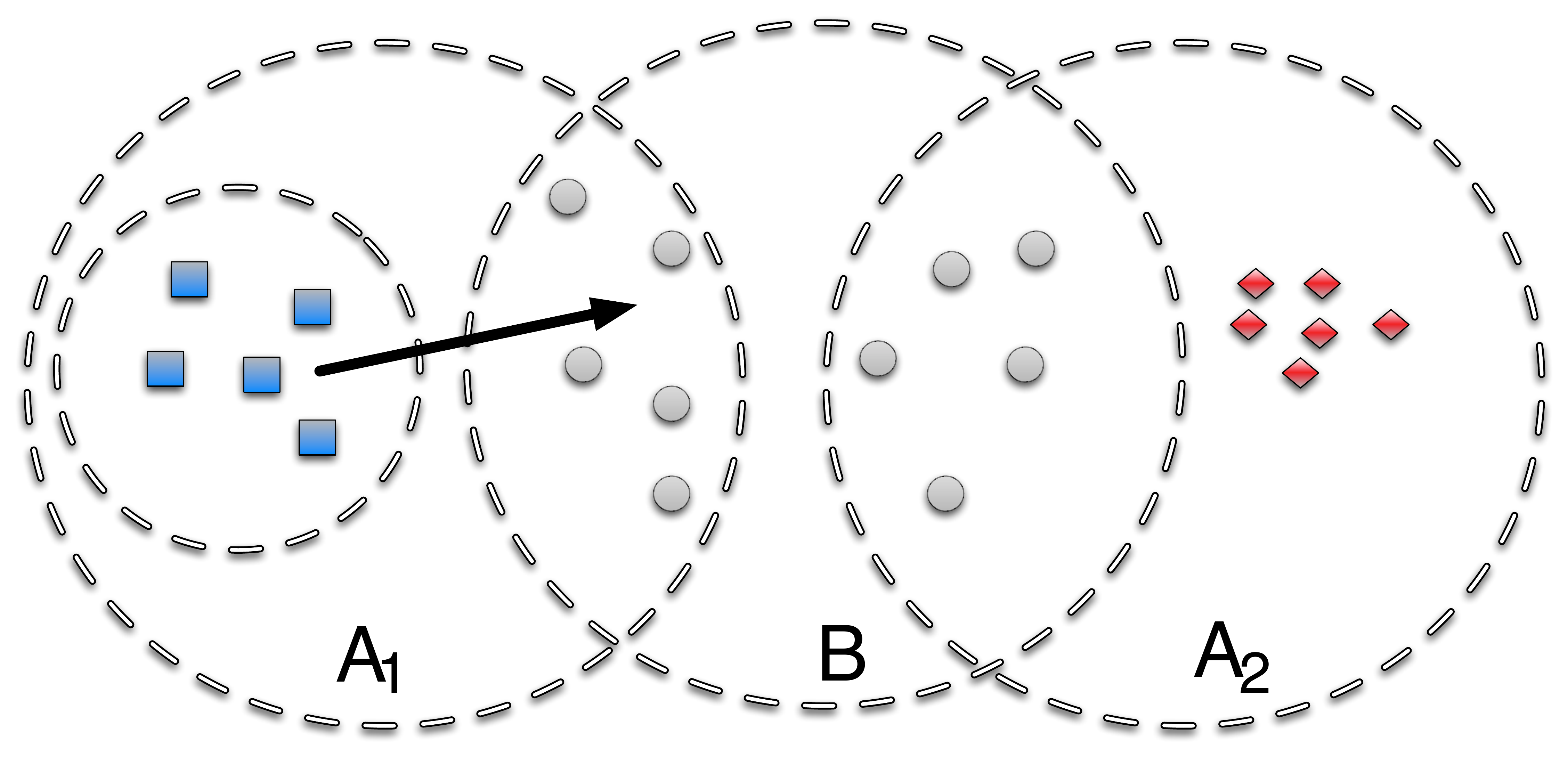}
   \caption{Area partitioning for attempting double spending.}
    \label{fig:dbsp}
\end{figure}

To complete a transaction in both components, each must contain enough users to verify the transaction and are at least $\mathit{aVd}$ apart. This provides a lower bound on the number of users that must be controlled to create a suitable cut. Suppose a malicious user manages to induce an artificial cut as shown in Figure \ref{fig:dbsp}. He partitions the users into two components, $A_1$ with $\mathcal{U}(A_1)$ users and $A_2$ with $\mathcal{U}(A_2)$ users by controlling users in region B. Let $A = A_1\setminus B$ and the number of the users in $A$ are $ |A| = \alpha \frac{|\mathcal{U}|}{2}$. Also, let the average distance between any pair of users within $A$ to be $\gamma \mathit{aVd}$. That is, 

\begin{equation}
\sum_{i,j| i,j \in A} \frac{|d_i - d_j|}{|A|/2(|A|-1)} = \gamma \mathit{aVd}. 
\end{equation}

Let the average distance of a user in region $B$ with a user in regions A or B to be $\zeta \mathit{aVd}$. The malicious user has to ensure that the average distance of any pair of users who agree the transaction is at least $\mathit{aVd}$. Let malicious user selects $\mathcal{M}$ users from region B for block creation with $\mathcal{U}(A_1)$. She needs to ensure:
\begin{align*}
\frac{\alpha \zeta |\mathcal{M}|\frac{|\mathcal{U}|}{2}  \mathit{aVd} + \zeta \frac{|\mathcal{M}|^2}{4} \mathit{aVd} + \gamma \frac{\alpha^2|\mathcal{U}|^2}{4} \mathit{aVd} }{\alpha |\mathcal{M}|\frac{|\mathcal{U}|}{2} + \frac{|\mathcal{M}|^2}{4} + \frac{\alpha^2|\mathcal{U}|^2}{4}} > \mathit{aVd} \Leftrightarrow \\
2\alpha |\mathcal{M}||\mathcal{U}|(\zeta-1) + |\mathcal{M}|^{2}(\zeta-1) + \alpha^2|\mathcal{U}|^2(\gamma -1) >0
\end{align*}

If the malicious user chooses to (i) increase $\zeta$ or (ii) decrease $\alpha$ and $\gamma$, the region B will enlarge and thus he will need to add more users into $\mathcal{M}$. Intuitively, to decrease $|\mathcal{M}|$ he needs, higher $\zeta$, or smaller values of $\alpha$ which in turn again need to control a bigger percentage of $\mathcal{U}$ and increase $|\mathcal{M}|$. 

\vspace{0.3 cm}
\fbox{\begin{minipage}{0.9\columnwidth}
\vspace{0.1 cm}
\textit{Example: } For $\alpha=1, \gamma=1/2$ and $\zeta=1.5$, $m$ has to control at least $|\mathcal{M}| \geq (\sqrt{2}-1) |\mathcal{U}|$ or over $41\%$ of the users.
\vspace{0.1 cm}\end{minipage}}\vspace{0.3 cm}

\begin{figure}[H]
\centering
    \includegraphics[width=0.7\columnwidth]{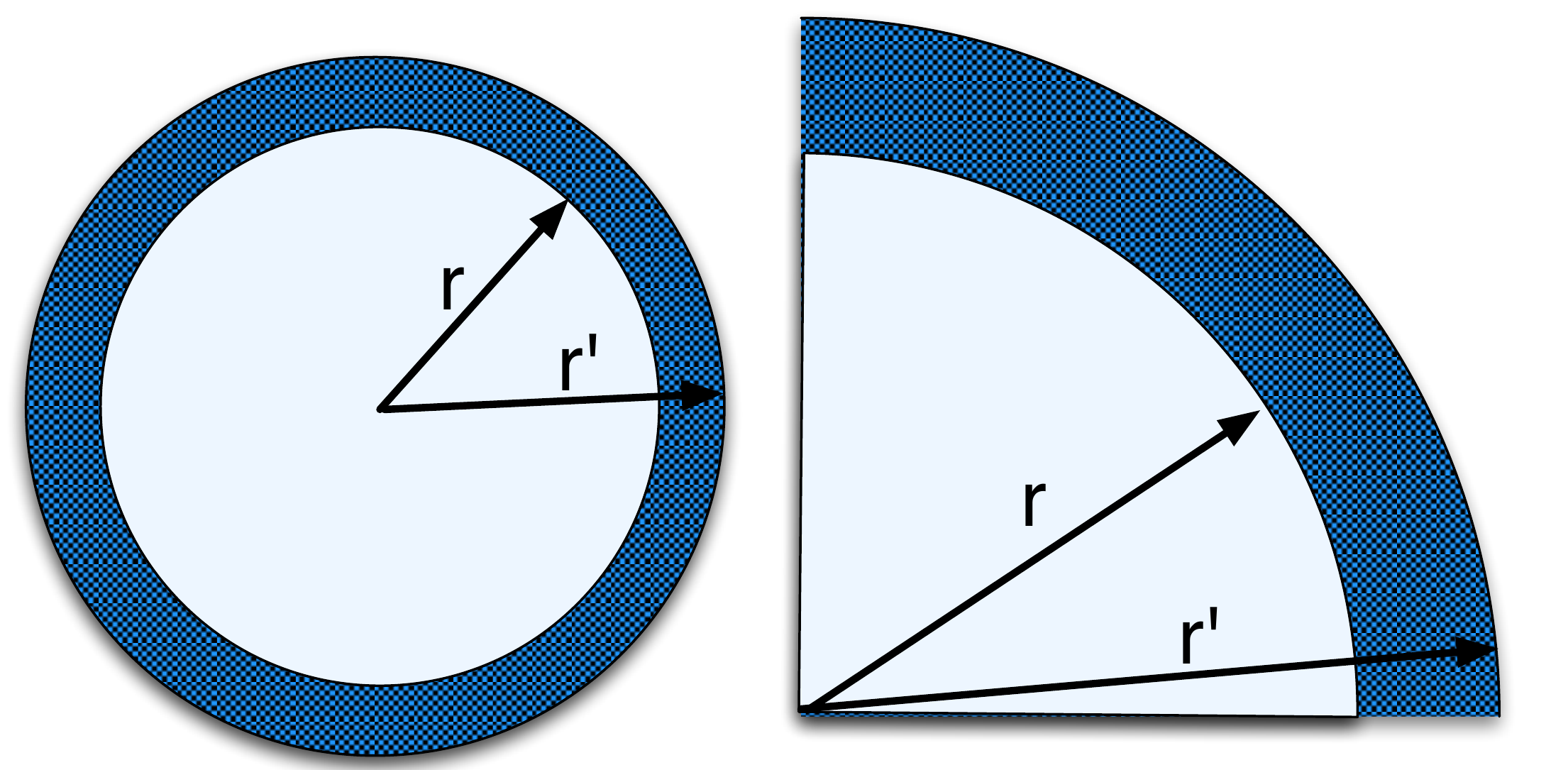}
   \caption{Area controling for attempting double spending.  }
    \label{fig:dsa}
\end{figure}

Note that the above double spending attack is one of the possible attack and an attacker can isolate a region $\mathcal{R}$ by controlling all the users in a specific region. If he manages to control all the users in the shaded region, as shown in Figure \ref{fig:dsa}, he can successfully create two virtual disconnected components in the connected graph. In order to block any message flow between the regions, $r'>r+2r_{cov}$ has to hold. That is, he needs to cover an area the size of at least 


\begin{eqnarray}
	\mathcal{R}= \pi\big(r'^2 - (r'-2r_{cov})^2\big) = 4\pi r_{cov}(r'-r_{cov})
\end{eqnarray}

The attacker can further optimize the attack by locating one of the two virtual components in a corner, so as to reduce the area he needs to control (Fig. \ref{fig:dsa}). In a region $A$, the average distance between any two users placed with a uniform distribution is $\tilde{d}r'$ (Under extensive simulations $\tilde{d} = 0.45$). In LocalCoin protocol, a transaction will be completed in $A$ if the average distance between any two users is at least $\mathit{aVd}$. That is, $\tilde{d}r'>\mathit{aVd}$. The attacker can reduce the area that he needs to control by trying to push $A$ further into a corner but he cannot reduce $r'<\frac{\mathit{aVd}}{\tilde{d}}$. Thus to create two successful transactions, one on $A$ and one in the remaining area $\mathcal{D}\setminus A$, before he gets detected, he needs to control all the users in an area the size of: 
\begin{eqnarray}
	\mathcal{R} &=&\frac{1}{4}\pi\big((r')^2-(r'-2r_{cov})^2\big) = \pi r_{cov}(r'-r_{cov})  \notag \\
	&>& \pi r_{cov} (\frac{\mathit{aVd}}{\tilde{d}}-r_{cov})
\end{eqnarray}

We assume that the users are uniformly distributed on a unit square.  The average distance between any two users on any disc of radius $r$ inside this unit square is $\bar{d}r$ (Under extensive simulations $\bar{d} = 0.903$). In LocalCoin protocol, we desire the information about each transaction to reach at least 50\% of the users at the time of accepting a transaction. Thus, $\pi r^2 > 0.5$ which translates to  $\mathit{aVd} > 0.36$.  A higher value of $\mathit{aVd}$ will slow down transactions' verification rate but it will increase security.  

\vspace{0.3 cm}
\fbox{\begin{minipage}{0.9\columnwidth}
\vspace{0.1 cm}
\textit{Example: } For $\mathit{aVd}=\frac{1}{3}$ and $r_{cov}=0.05$, the malicious user needs to control a region which is  $0.1085$ of $\mathcal{D}$. As all the users are uniformly distributed on $\mathcal{D}$, it amounts to controlling 10.85\% of $\mathcal{U}$. 
\vspace{0.1 cm}\end{minipage}}\vspace{0.3 cm}

\begin{theorem}\label{thm:virtual_attack}
To be able to double spend a LocalCoin by a virtual cut, an attacker needs to control at least $\mathcal{R} > \pi r_{cov} (\frac{\mathit{aVd}}{\tilde{d}}-r_{cov})$ fraction of the users, under the assumption that all the users are uniformly distributed and are static.
\end{theorem}

\subsubsection{Virtual-cut Attack: An active double spending attack in dynamic networks}\label{sssec:vc_dn}
In reality, even if the attacker colludes with $\mathcal{M}: |\mathcal{M}| > \frac{|\mathcal{R}|}{|\mathcal{D}|}$ other users and places them appropriately to create two virtual components in the network. The remaining $\mathcal{U} \setminus \mathcal{M}$ users may be moving dynamically. So at the time when he plants a double spending attack,  for the attack to be successful, none of these $|\mathcal{U}|-|\mathcal{M}|$ users should be placed in $\mathcal{R}$.  Thus the probability that such the double spend attack by controlling large number of users to be successful is 
\begin{equation}
\frac{|\mathcal{R}|}{|\mathcal{D}|}^{(|\mathcal{U}|-|\mathcal{M}|)}
\end{equation}

\vspace{0.3 cm}
\fbox{\begin{minipage}{0.9\columnwidth}
\vspace{0.1 cm}
\textit{Example: } For $\frac{|\mathcal{R}|}{|\mathcal{D}|} \approx 10\% $ and $|\mathcal{U}|-|\mathcal{M}| \geq 100$, the probability of a successful attack is at most $10^{-6}$.
\vspace{0.1 cm}\end{minipage}}\vspace{0.3 cm}
The assumptions made on the analysis of \textit{LocalCoin} are discussed in Section \ref{sec:assumptions}.

\begin{figure}[t]
    \centering
\begin{subfigure}{\columnwidth}
\centering
   \includegraphics[height=0.8\columnwidth, angle=270]{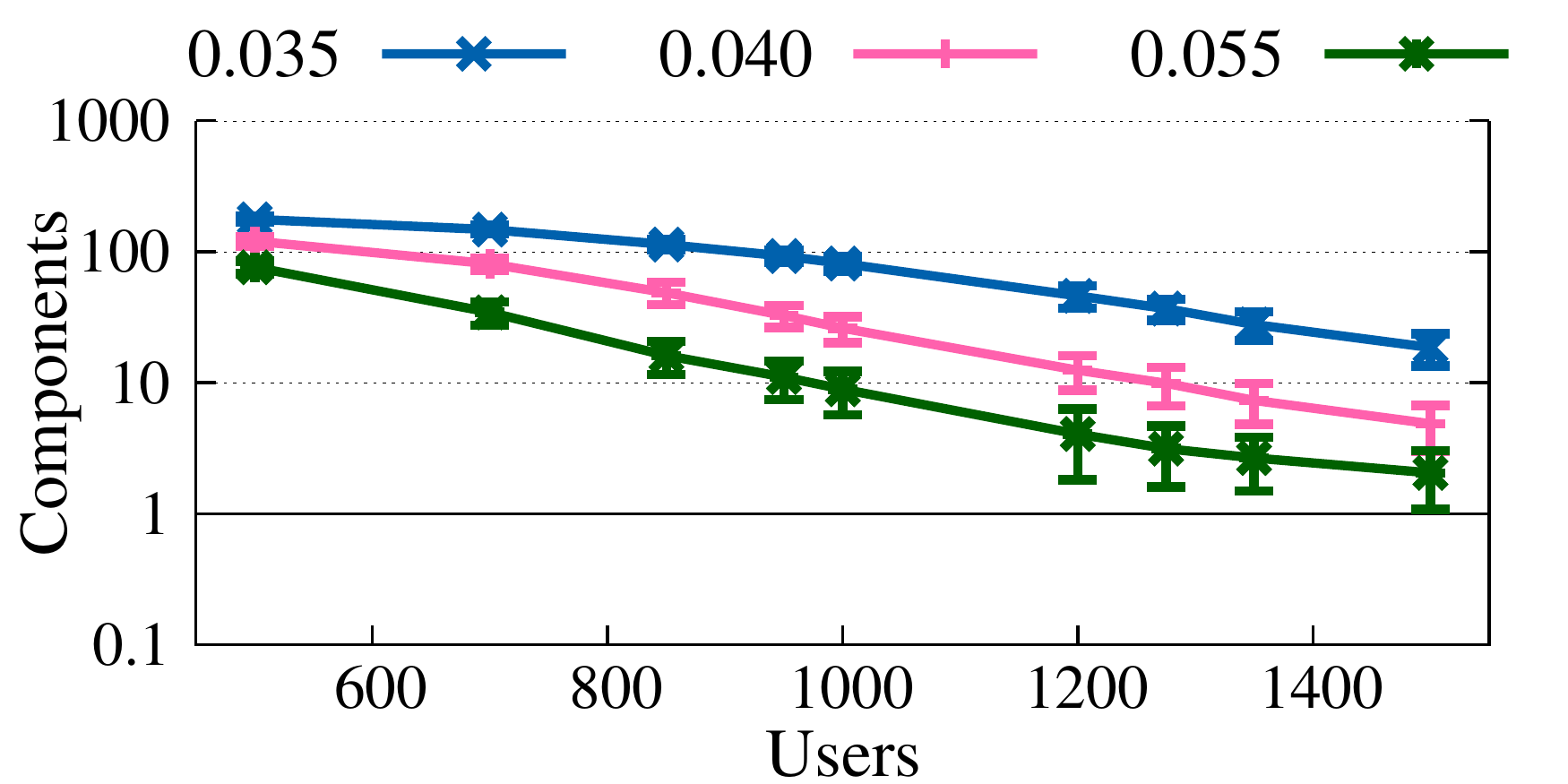}
   \includegraphics[height=0.8\columnwidth, angle=270]{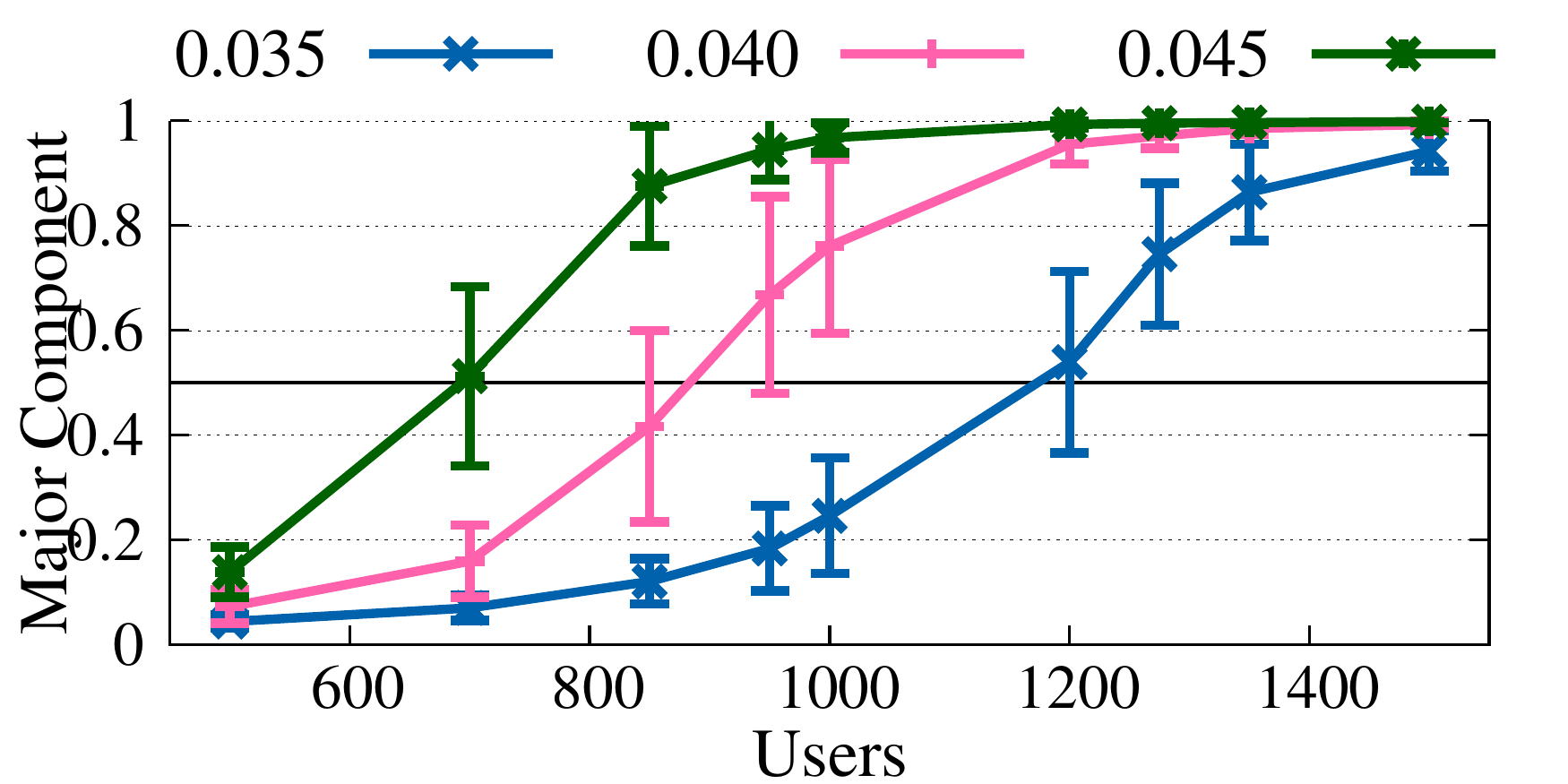}
    \caption{Different $|\mathcal{U}|$}
    \label{fig:users}
\end{subfigure}
\begin{subfigure}{\columnwidth}
\centering
    \includegraphics[height=0.8\columnwidth, angle=270]{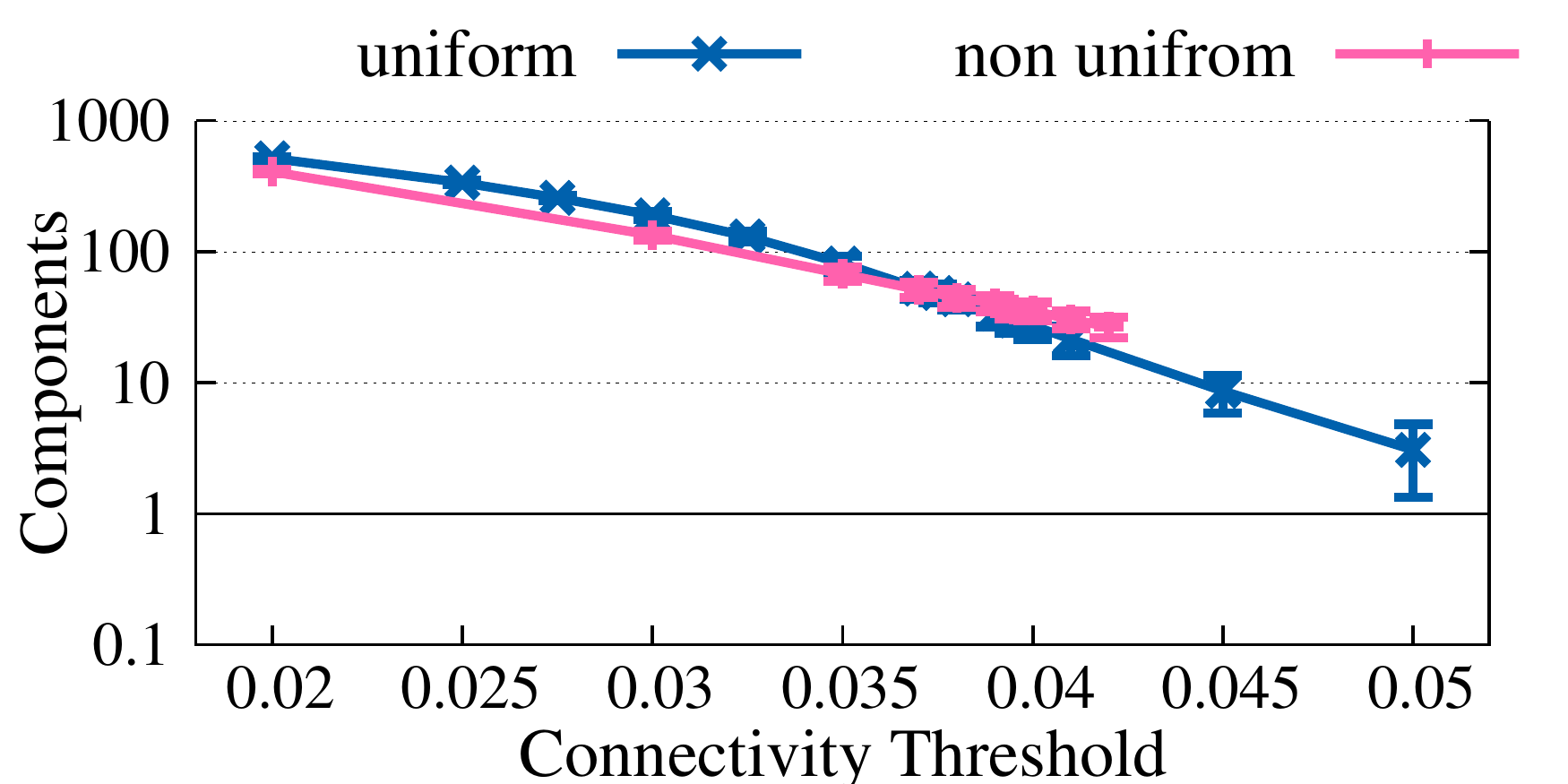}
   \includegraphics[height=0.8\columnwidth, angle=270]{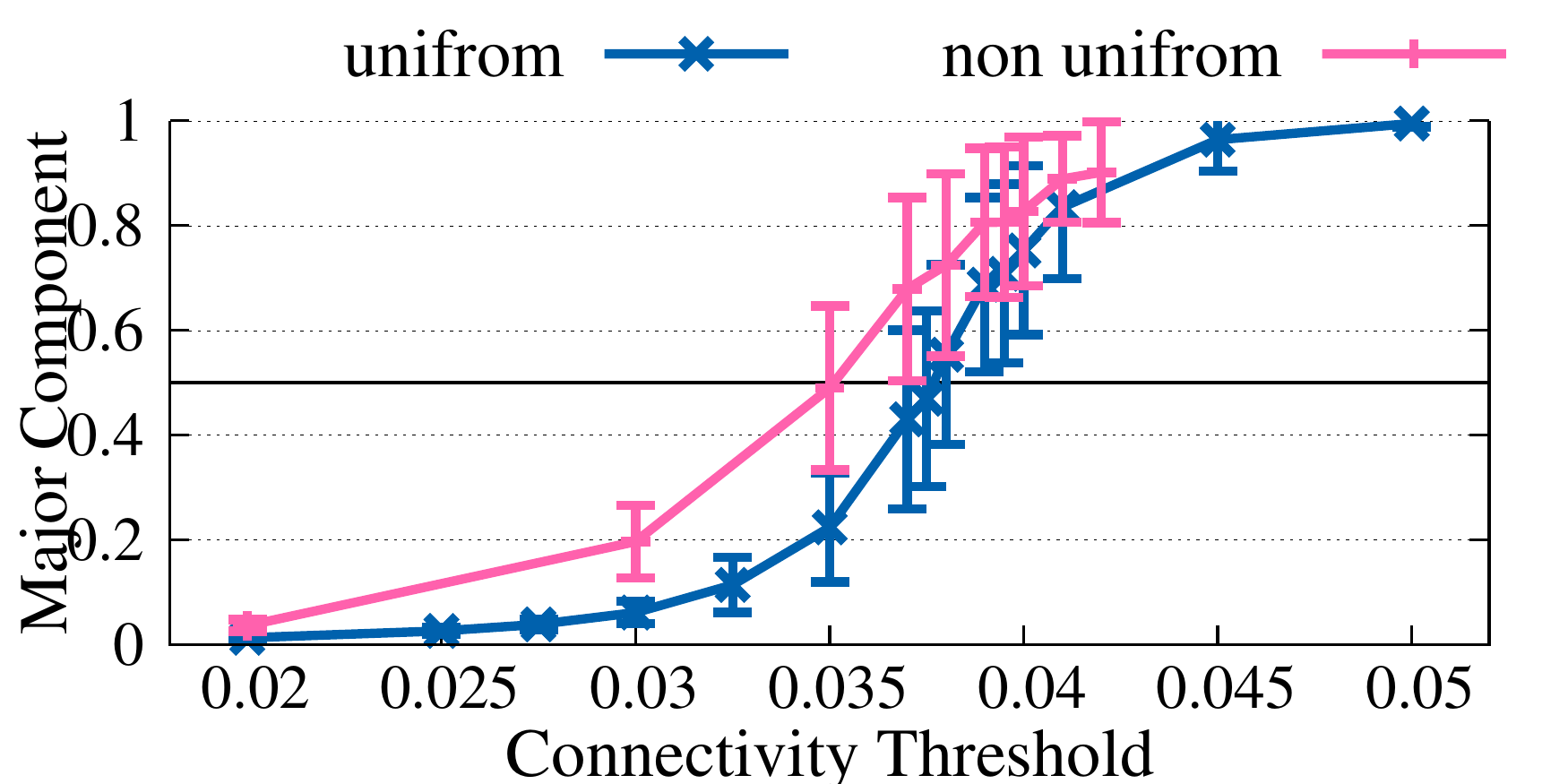}
   \caption{Different $r_{cov}$}
   \label{fig:thresh}
\end{subfigure}
    \caption{The connected components and the fraction of the users in the major component in a $[0,1]\times[0,1]$ area.}
    \label{fig:basics}
\end{figure}

\begin{figure*}[t]
    \centering
    \begin{subfigure}{0.65\columnwidth}
    	\centering
    	\includegraphics[width=0.7\columnwidth,angle=270]{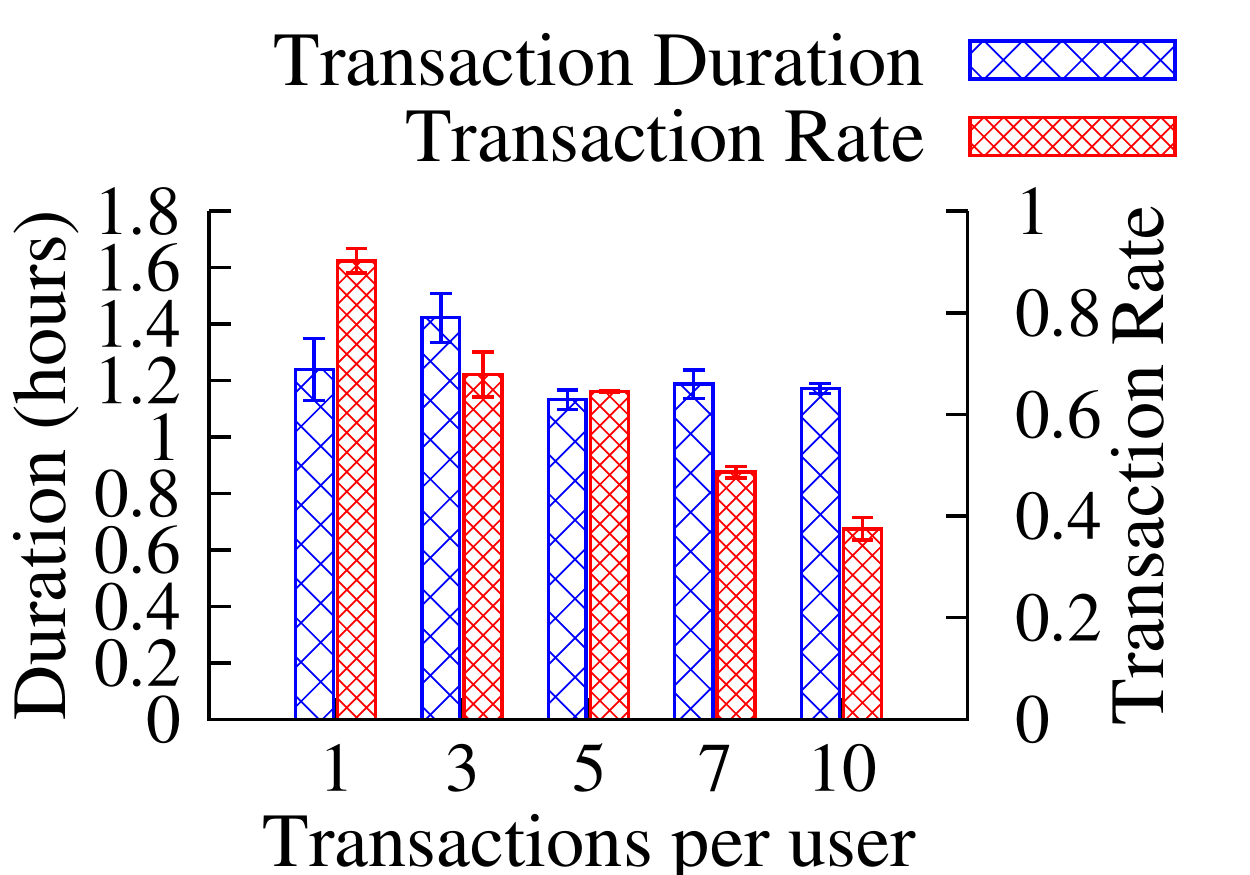}
   		\caption{Infocom 05}
        \label{fig:inf05normal}
	\end{subfigure}
    \begin{subfigure}{0.65\columnwidth}
    	\centering
    	\includegraphics[width=0.7\columnwidth,angle=270]{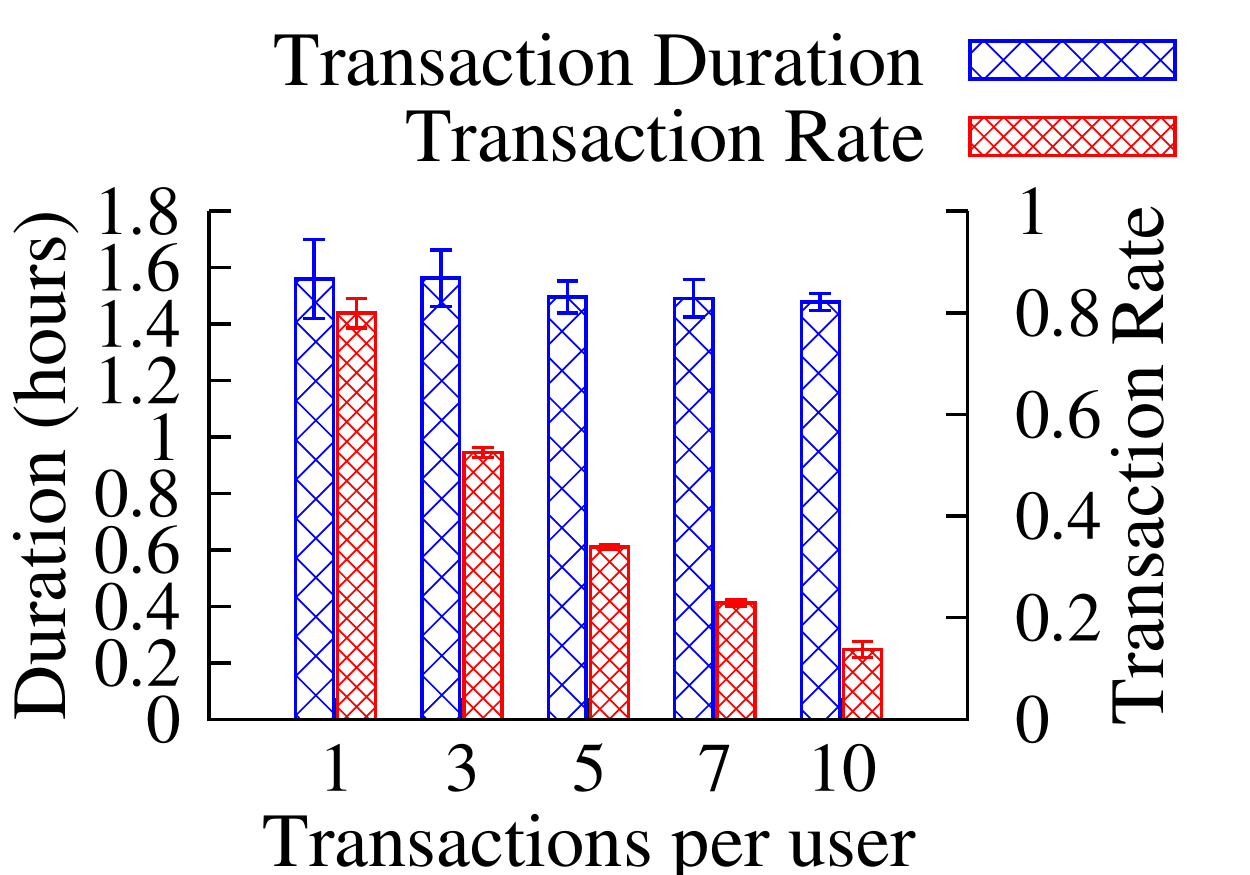}
   		\caption{Infocom 06}
        \label{fig:inf06normal}
	\end{subfigure}
    \begin{subfigure}{0.65\columnwidth}
    	\centering
    	\includegraphics[width=0.7\columnwidth,angle=270]{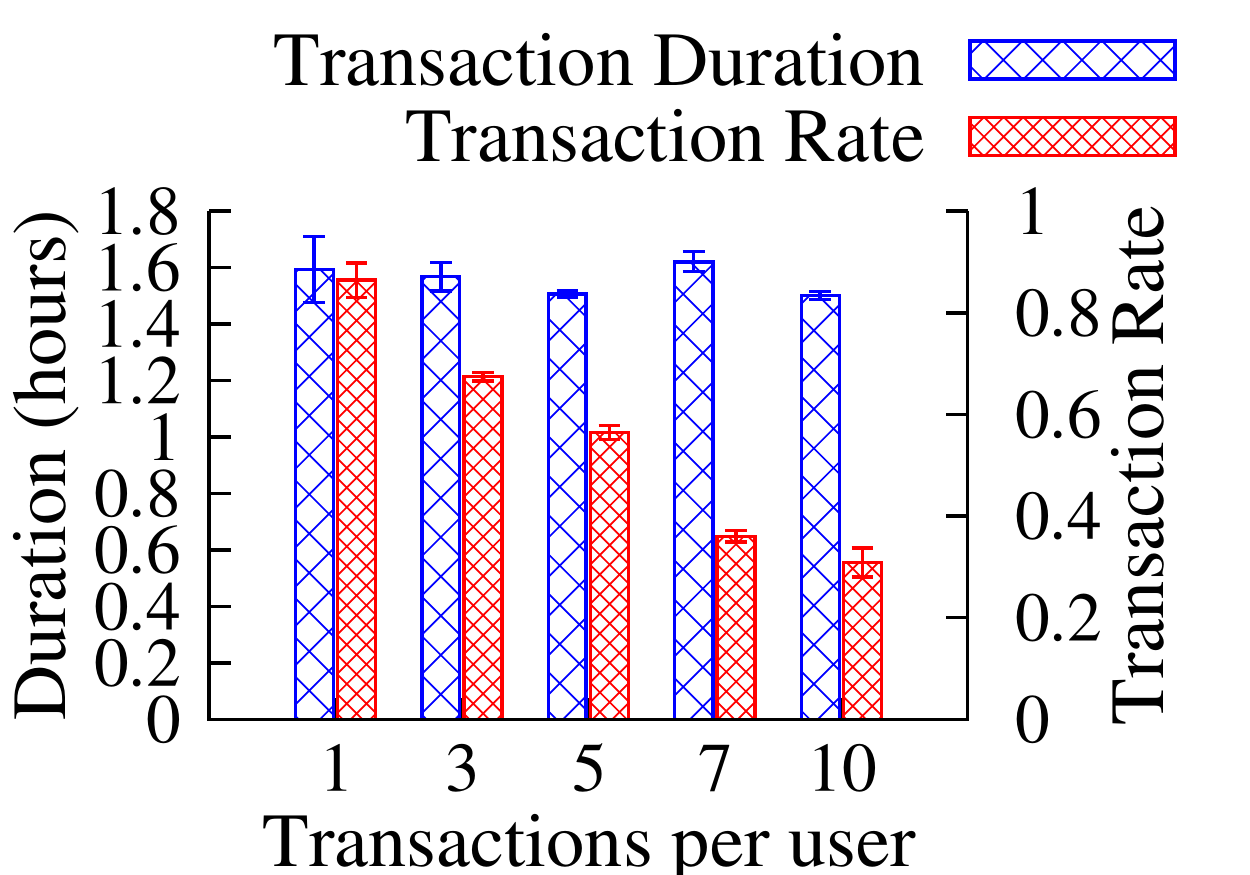}
        \caption{Humanet}
   		\label{fig:humnormal}
	\end{subfigure}
    \begin{subfigure}{0.65\columnwidth}
    	\centering
    	\includegraphics[width=0.7\columnwidth,angle=270]{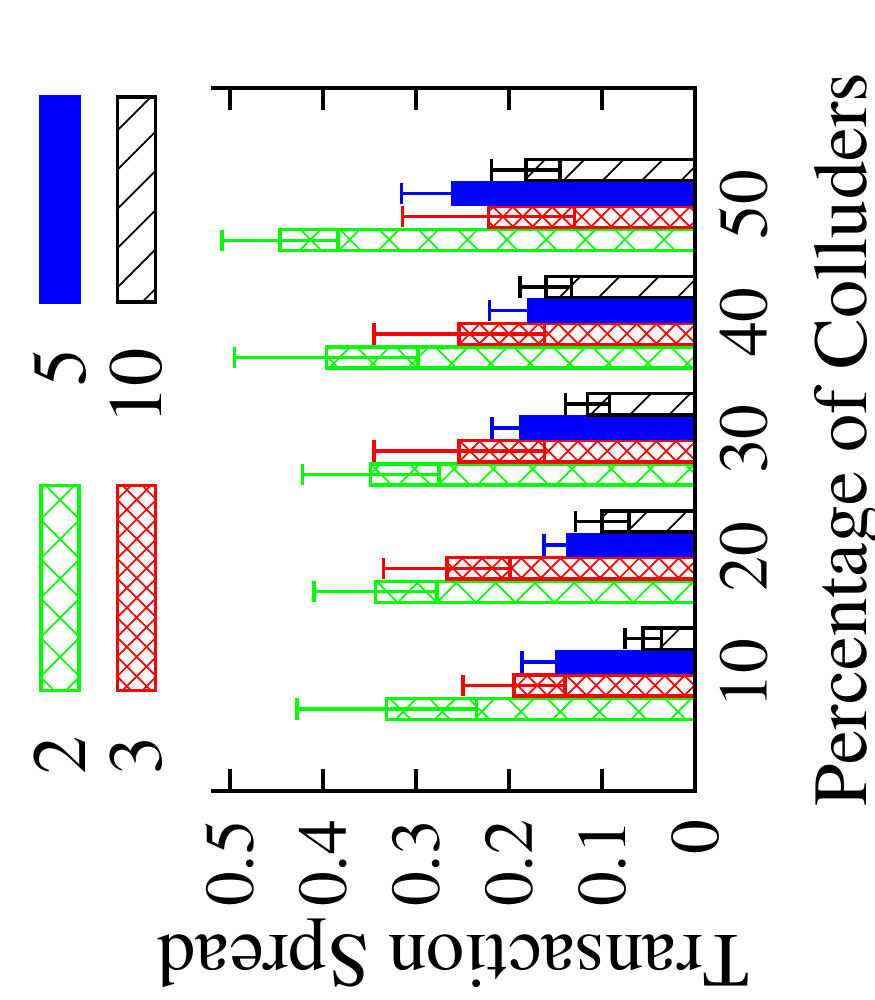}
   		\caption{Infocom 05}
        \label{fig:inf05comp}
	\end{subfigure}
    \begin{subfigure}{0.65\columnwidth}
    	\centering
    	\includegraphics[width=0.7\columnwidth,angle=270]{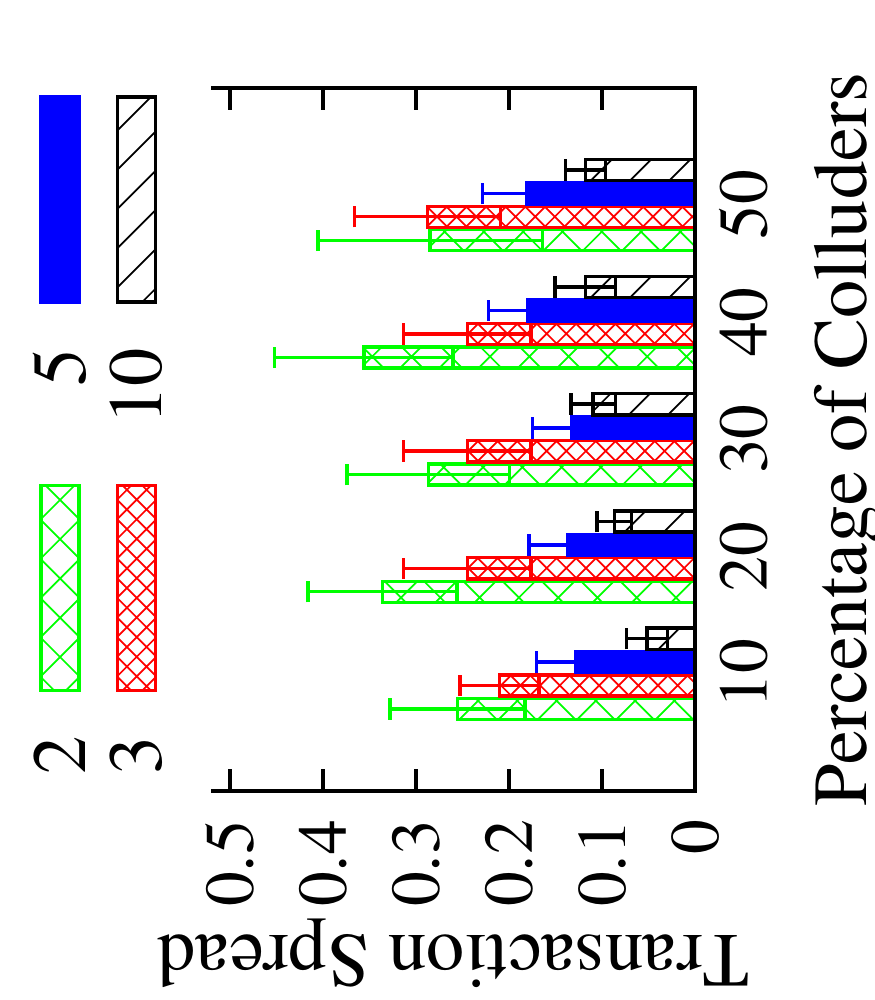}
       \caption{Infocom 06}
 		\label{fig:inf06comp}
	\end{subfigure}
        \begin{subfigure}{0.65\columnwidth}
    	\centering
    	\includegraphics[width=0.7\columnwidth,angle=270]{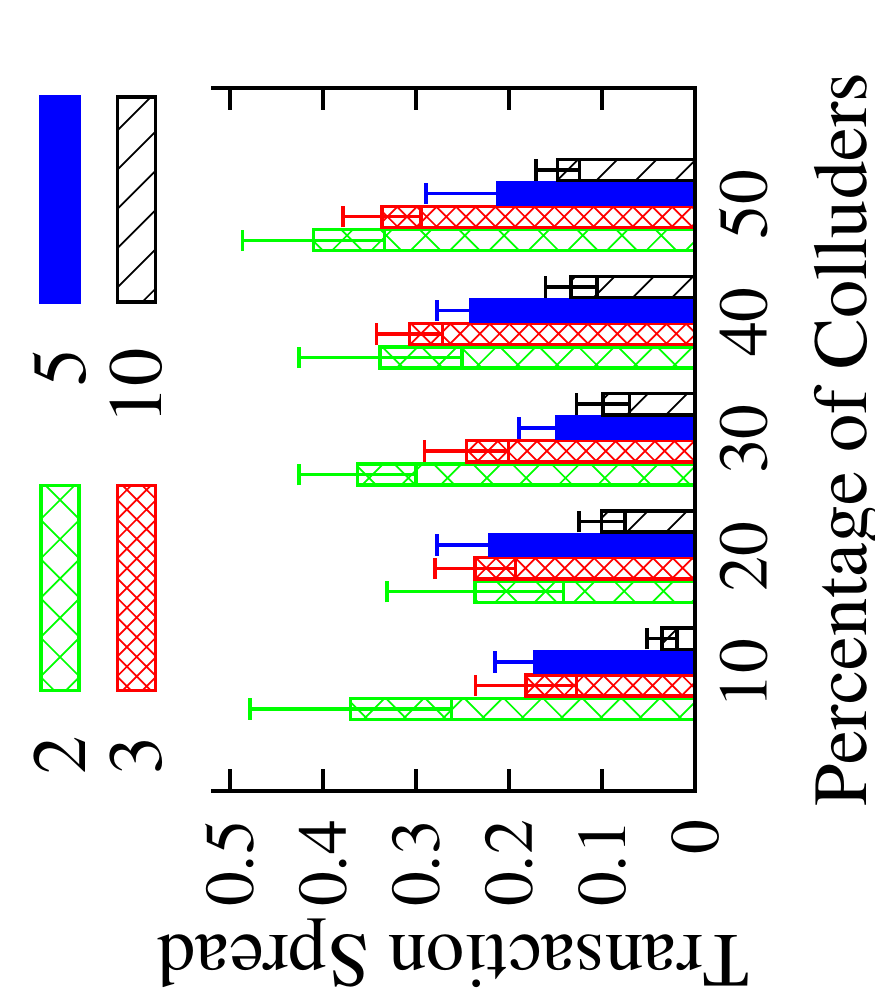}
   		\caption{Humanet}
        \label{fig:humcomp}
	\end{subfigure}
    \caption{Analysis of LocalCoin using users' mobility from Infocom 05, Infocom 06 and Humanet traces.\vspace{-0.4cm}}
    \label{fig:traces}
\end{figure*}

\section{Performance Evaluation of LocalCoin }\label{sec:simulations}

We conduct a set of simulations in static and dynamic graphs. The static random geometric graphs are produced with MATLAB (Section \ref{sec:matlabsims}). We implement an event driven simulator in JAVA for dynamic graphs using real mobility traces (Section \ref{sec:javasims}) and in order to scale up the number of the mobile users, we use the ONE simulator \cite{one} (Section \ref{sec:onesims}). After proving, in Section \ref{sec:analysis}, that double spending is improbable in fully connected mobile ad-hoc networks, we investigate scenarios where the users are not fully connected in order to depict the robustness of LocalCoin.

\subsection{Evaluation with RGGs} \label{sec:matlabsims}
In the static analysis, we focus on the characteristics of the produced RGGs and their affect on LocalCoin. The simulations on static graphs are important because in the case where users are not moving, a malicious user $m$ has the highest chances to successfully double spend some localcoins. We distribute uniformly the users $\mathcal{U}$ in a $[0,1]\times[0,1]$ area and we study the effect of $|\mathcal{U}|$ and $r_{cov}$ on the number of connected components and the fraction of users in the largest connected component. 
\begin{figure}[t]
    \centering
   \includegraphics[width=0.8\columnwidth]{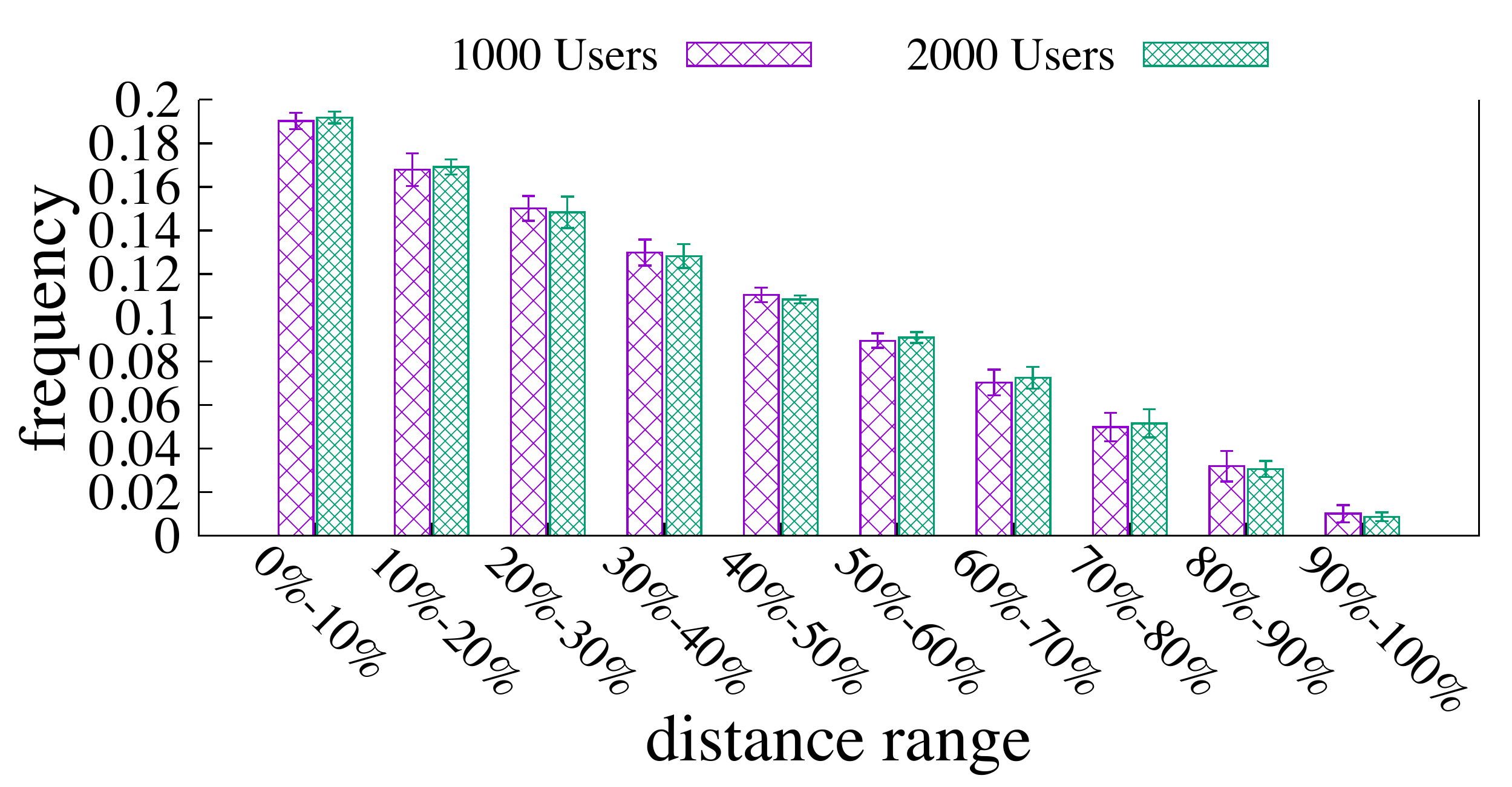}
    \caption{Average distance between a randomly selected user with all the other users. \vspace{-0.3cm}}
    \label{fig:avgDist}
\end{figure}
Figure \ref{fig:users} shows how $|\mathcal{U}|$ affects the number of connected components and the size of the major connected component for three different values $r_{cov}$. 
\begin{wrapfigure}[11]{l}{.35\columnwidth}
    \centering
   \includegraphics[width=0.3\columnwidth]{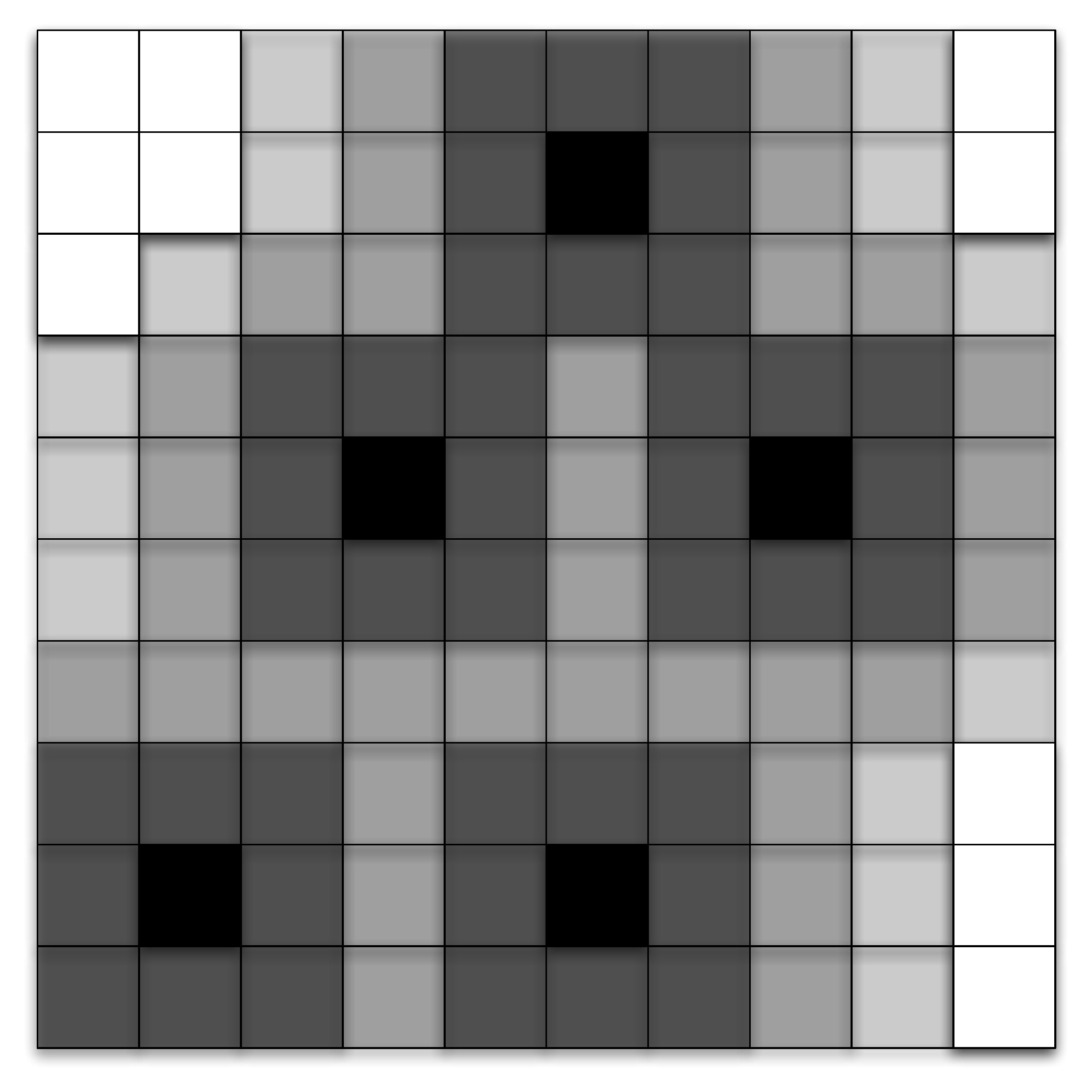}
    \caption{Non-uniform distribution of users.}
    \label{fig:grid}
\end{wrapfigure}
Figure \ref{fig:thresh} shows how $r_{cov}$ affects the aforementioned quantities for two different user placements. 
Given that the uniform distribution of $\mathcal{U}$ is not a realistic case, we split the examined area into a 10 by 10 grid of equally spaced 100 cells,  where the number of users in each cell is determined by a Poisson distribution with different parameter. Figure \ref{fig:grid} depicts the used grid and the darkness of each cell depicts its populatiry. 
The average number of the total users is still 1000. In this non uniform case, the major component is formed for smaller values of $r_{cov}$.
In the case of static graphs, double spending is possible when the number of connected components is more than 1, $\mathit{mVu}$ is smaller than the number of the users in the components selected to double spend and $\mathit{aVd}$ is small enough to apply for the users in each component. This requires at least two large components with roughly equal size. From our simulations, for $|\mathcal{U}|=1000$ and $r_{cov}=0.5$, the major component contains more than 90\% of the users. As only one big component is getting formed the information about each transaction will spread in the network very easily. 

Figure \ref{fig:avgDist} presents the distribution of the users in the examined area and it is useful to understand the impact of the constraint of the average distance between the users who verify the creation of a new block on the spread of the block creation process in the whole area. In order to produce the figure, we placed randomly 1000 and 2000 users and we selected randomly one of them and we measure the distance of all the other users with the selected one. Figure \ref{fig:avgDist} shows that even if the imposed threshold in the creation of a block is less that 30\% of the maximum possible value, more than 50\% of the users will be informed about the creation process. 
In order to examine LocalCoin in more realistic cases, we consider mobile users in the next two subsections.

\subsection{Evaluation with Mobility Traces}\label{sec:javasims} 

We implement an event-driven simulator in Java in order to depict the performance of \textit{LocalCoin}. We used three datasets, Infocom'05 and Infocom'06 from the Haggle project \cite{cambridge-haggle-2006-01-31} and Humanet \cite{tecnalia-humanet-2012-06-12}, which contain user mobility traces in different environments. The duration of the simulation is one day. We select the first day of the first two datasets, while Humanet is one day long. We considered all the mobile users, which are 41, 78 and 56 respectively. 

\begin{figure}
	\centering
    	\includegraphics[height=0.6\columnwidth,angle=270]{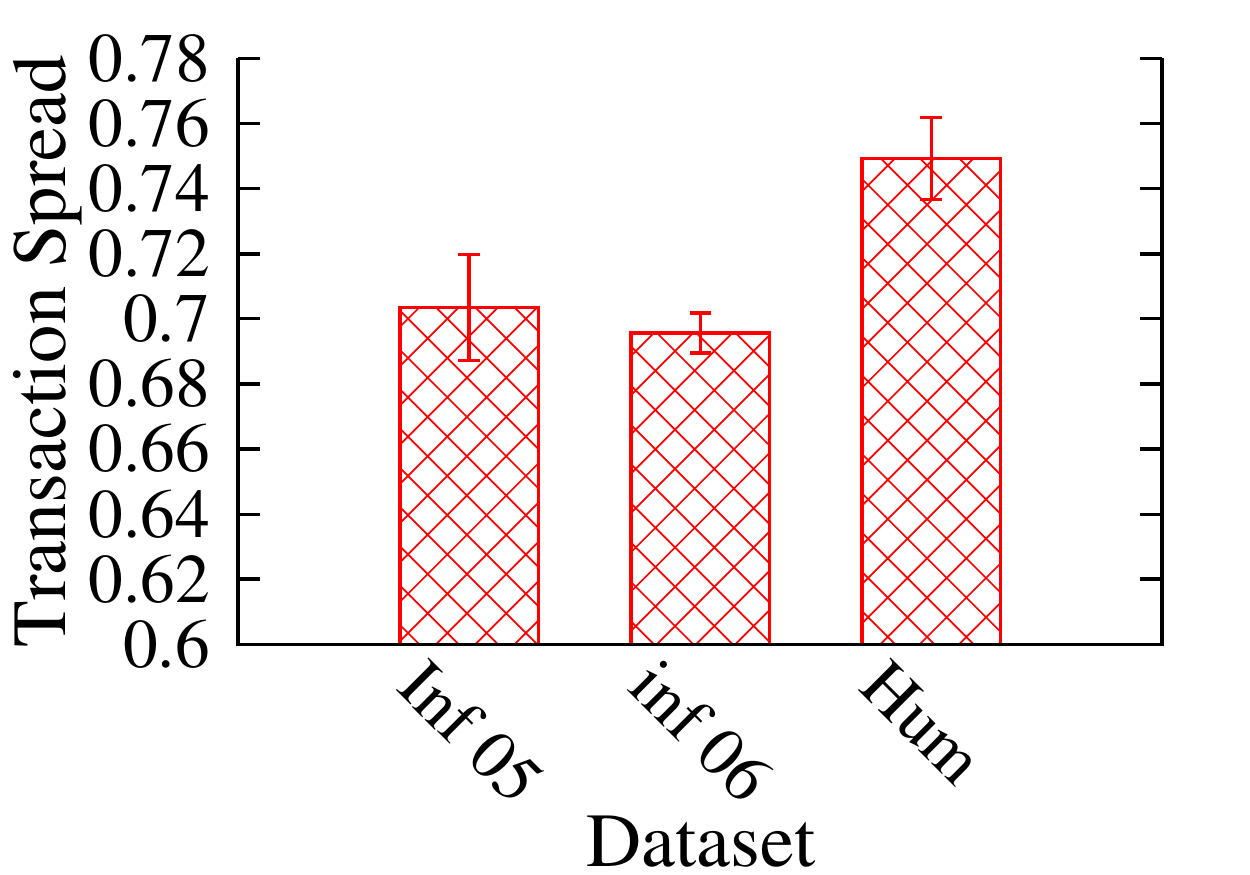}
   	\caption{The fraction of the transactions that were delivered to the destination.\vspace{-0.4cm}}
        \label{fig:compareAll}
\end{figure}

We introduce the datasets using the concepts of \textbf{Transaction Rate} and \textbf{Transaction Spread}. We define transaction rate as the fraction of the completed transactions and the transaction spread as the average fraction of the users that have stored the transaction. Figures \ref{fig:inf05normal}, \ref{fig:inf06normal} and \ref{fig:humnormal} show the average time needed for one transaction to reach its destination and the transaction rate for different number of transactions per user. The receiver and the time of the transaction occurrence are generated uniformly between the users and the day. Figure \ref{fig:compareAll}, illustrates the transaction spread in the case where each user initiates one transaction at the beginning of the day. Note that all datasets are sparse with small number of users and hence, the transaction spread is slow resulting into the small values of transaction rate and transaction spread. This set of figures is needed to explain and understand the results produced by the simulation of a malicious user in these three traces.

\begin{figure}[t]
    	\centering
    	\includegraphics[height=0.6\columnwidth,angle=270]{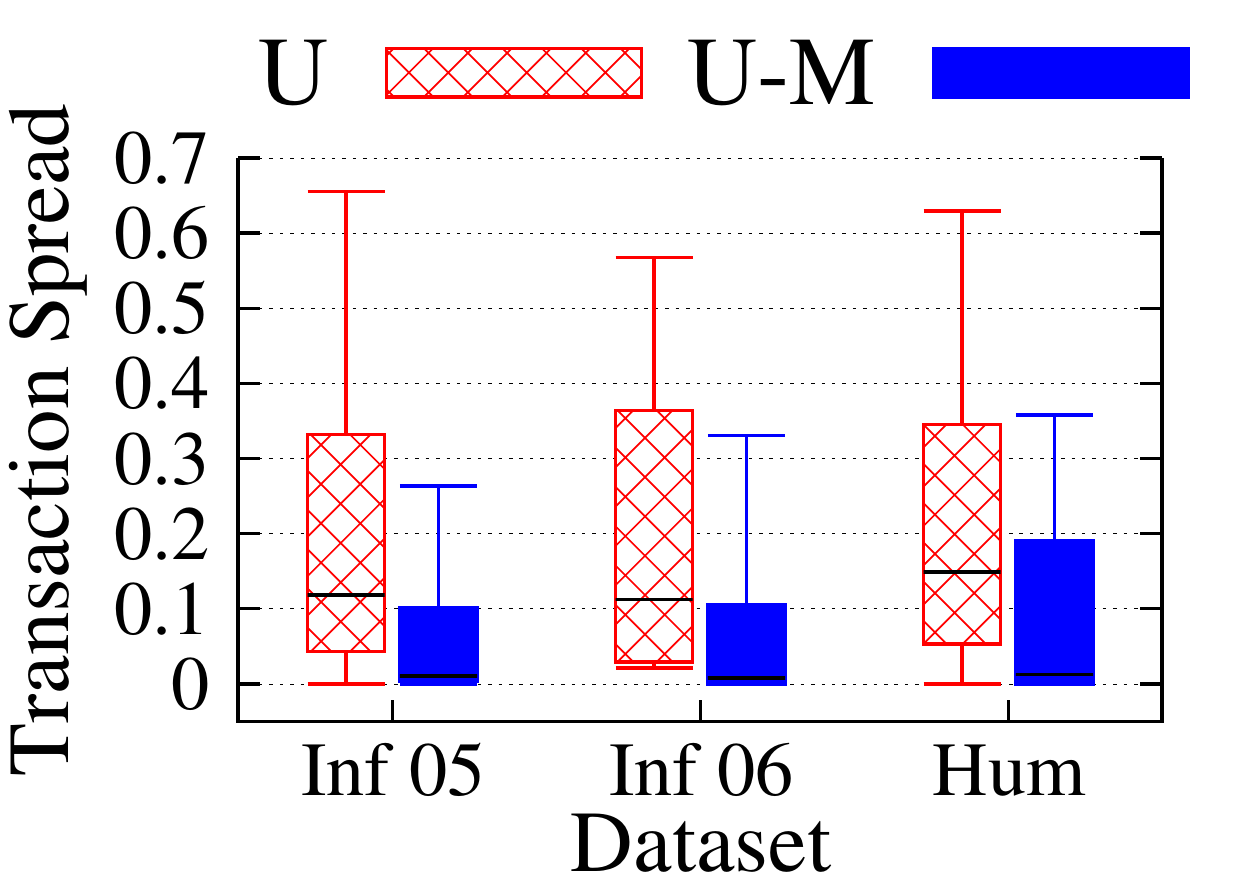}
	\caption{Spread of Fake Transactions.\vspace{-0.4cm}}
        \label{fig:badSpread}
\end{figure}

Next, we examine the chances a malicious user ($m$) has to deliver multiple transactions with the same input (\textbf{fake transactions}) to more than one users. $m$ tries to double spend by making at least two of the receivers of his fake transactions to accept them. However, double spending will not be successful before the creation of two blocks that contain these fake transactions, which is not possible if $\mathit{aVd}$ is large enough. To simulate a double spending attack: $m$ creates 2, 3, 5 or 10  fake transactions.
Figures \ref{fig:inf05comp}, \ref{fig:inf06comp} and \ref{fig:humcomp} show the average transaction spread of the fake transactions for variable number of colluders ($\mathcal{M}$). Multiple copies of the same transaction decrease the average spread of the fake transaction because the normal users ($\mathcal{U} \setminus \mathcal{M}$) receive at least two fake transactions with higher probability. Furthermore, most of these duplicates are stored by the colluders and not by the normal users. Figure \ref{fig:badSpread} shows the spread of the duplicates in $\mathcal{U}$ and in $\mathcal{U} \setminus \mathcal{M}$ for the case of 5 fake transactions and $|\mathcal{M}| = 0.5 |\mathcal{U}|$. On average, less than 2\% of the normal users receive the fake transactions. 
\begin{figure}[t!]
	\centering
	\includegraphics[height=0.6\columnwidth,angle=270]{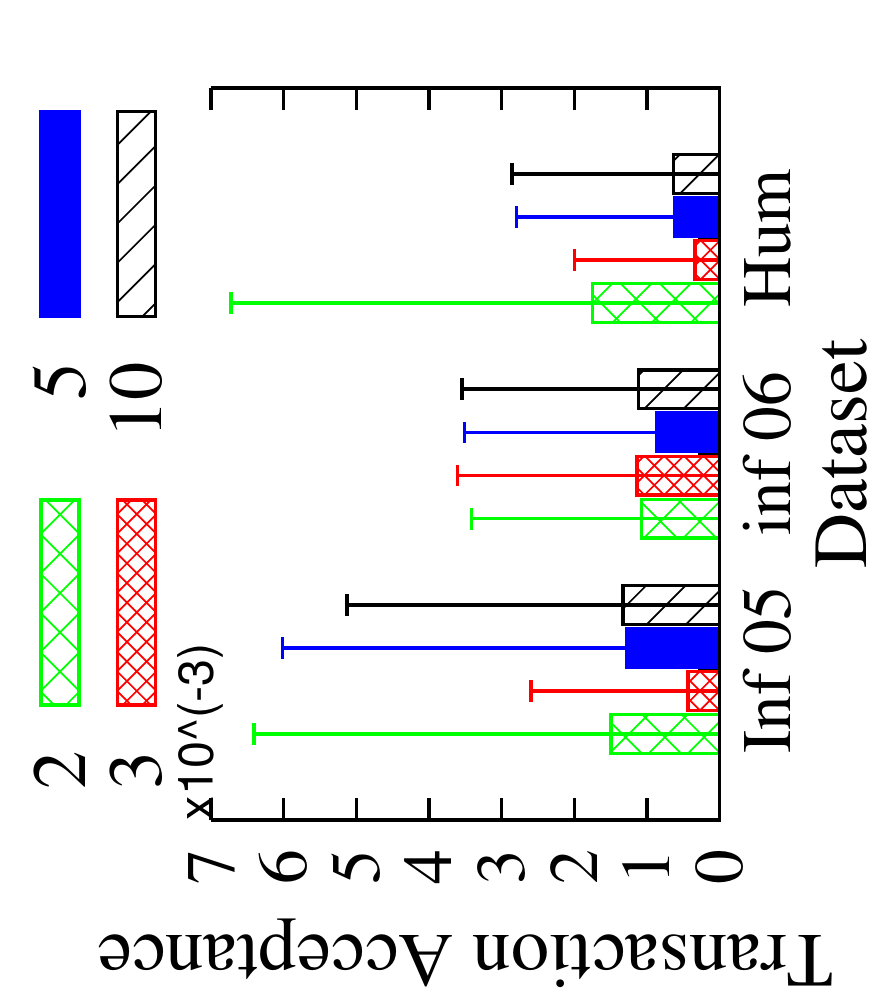}
	\caption{Accepted Fake Transactions.\vspace{-0.4cm}}
	\label{fig:fakereceived}
\end{figure}
Only the first copy has spread like a normal transaction while the spread of others is decreasing dramatically since normal users are familiar with the first ones.
Figure \ref{fig:fakereceived} shows the probability of at least one of the receivers of the fake transactions to accept the transaction. For analysis purpose, we make it easier for $m$ by using $\mathit{mTr}=0$. Even in this setting, the chance of $m$ being successful in making a receiver to accept a fake transaction is $<1$\%. Note that successful transaction does not mean that the transaction is accepted in blockchain but it is pending.

\subsection{Evaluation with ONE simulator}\label{sec:onesims} 
We implement LocalCoin on ONE simulator to scrutinize its behaviour on a larger scale and examine how users' speed and coverage radius affect the chances of a malicious user to double spend. We consider a 4$km^2$ area in the center of a Metropolitan city and a $|\mathcal{U}| = 1000$ mobile users. As a mobility pattern we use shortest path map based movement.

\begin{figure}[t]
\centering
\begin{subfigure}{0.8\columnwidth}
    \centering
    \includegraphics[height=\columnwidth, angle=270]{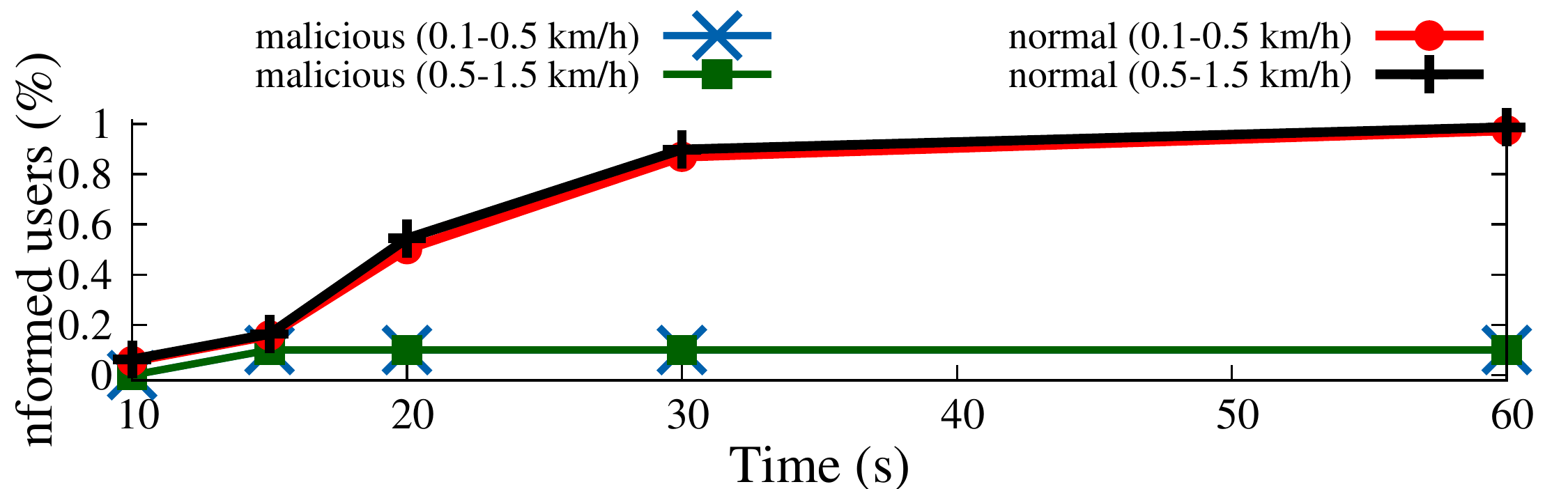}
   \caption{coverage radius = 100 meters}
   \label{fig:howMuchtime}
\end{subfigure}
\begin{subfigure}{0.8\columnwidth}
\centering
    \includegraphics[height=\columnwidth,angle=270]{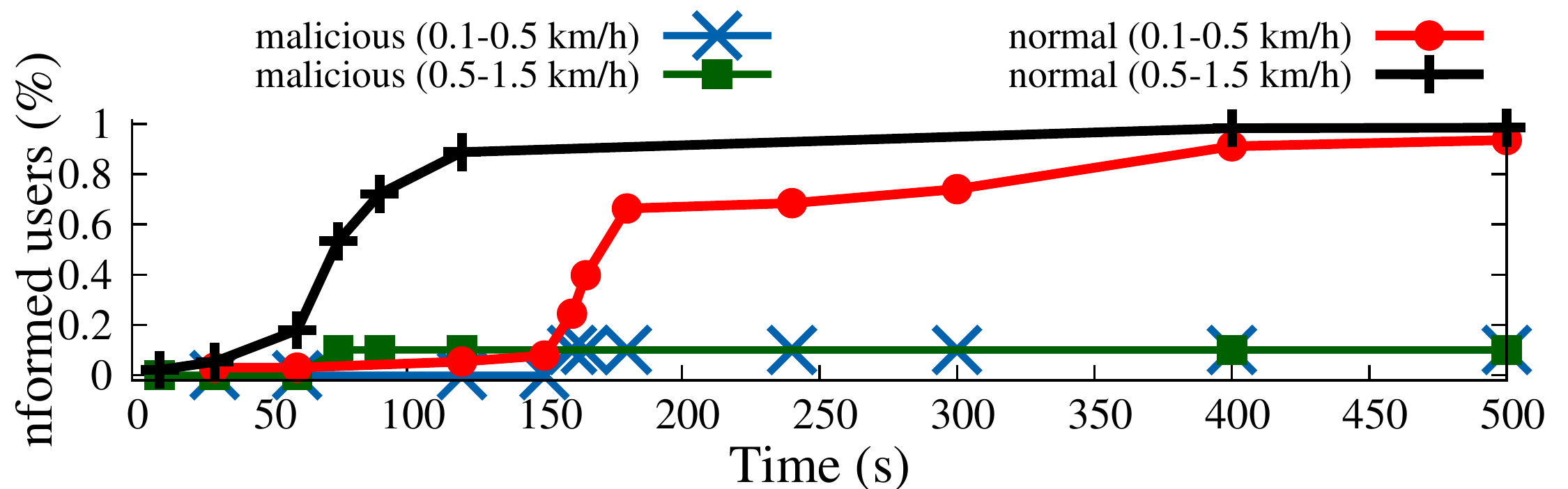}
    \caption{coverage radius = 50 meters}
    \label{fig:howMuchtimeSlower}
\end{subfigure}
\begin{subfigure}{0.8\columnwidth}
\centering
    \includegraphics[height=\columnwidth,angle=270]{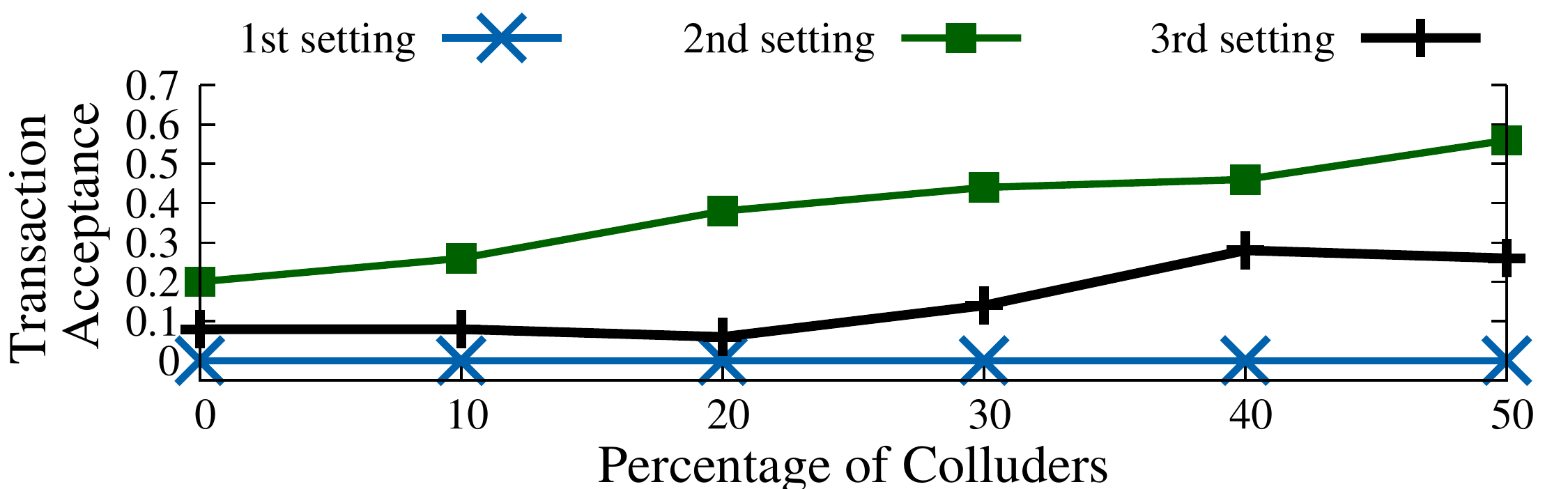}
    \caption{Prob. of Accepting a Fake Transaction}
    \label{fig:prob}
\end{subfigure}
    \caption{Analysis of LocalCoin protocol using ONE~\cite{one}.\vspace{-0.4cm}}
    \label{fig:theone}
\end{figure}

Figures \ref{fig:howMuchtime} and \ref{fig:howMuchtimeSlower} demonstrate the spread of a normal transaction. 
In Figure \ref{fig:howMuchtime}, the coverage radius of every broadcast is 100 meters while in Figure \ref{fig:howMuchtimeSlower} it is 50 meters\footnote{Wifi-direct supports a coverage radius of 200 meters but we expect it to drop radically in crowded urban areas.}.
Figure \ref{fig:howMuchtime} shows that even if a malicious user has $|\mathcal{M}| = 100$ colluders to forward his fake transaction he does not have enough time to create a second fake transaction because in less than a minute almost all the users in the area will be informed of her transaction. The walking speed of the users has little influence on this case. However, if the coverage radius is 50 meters (Figure \ref{fig:howMuchtimeSlower}), the walking speed matters. In case of the slow movement (0.1-0.5 km/h) it takes around 3 minutes for one broadcasted message to reach more than half of the users while in the case of normal walking (0.5-1.5 km/h) it takes less than 1 minute. 

In order to examine more thoroughly the case when cheating is still possible, we consider the case of normally walking users (0.5 - 1.5 km/h) and coverage radius of 50 meters. A malicious user $m$ creates two fake transactions and randomly selects the receivers of them. We examine three settings that differ in the time between the creation of the fake transactions. Figure \ref{fig:prob} depicts how the fraction of the colluders $\frac{|\mathcal{M}|}{|\mathcal{U}|}$ affects $m$'s chances to deliver the two fake transactions.
In the first setting, $m$ is not able to deliver 2 fake transactions, regardless of $|\mathcal{M}|$, because the time difference in the creation of the 2 them is only 10 seconds and $m$ is in contact with the same users. In the second setting the time difference is 1 minute and $m$ is able to successfully deliver two fake transactions but this happens either because he selects two users who have not met yet or have not met the same other users. In the third setting, $m$ initiates the two transactions with 2 minutes difference. In this setting, $m$, delays the creation of the second fake transaction as much as possible in order to move as far as possible from the users that have the first transaction. Then, he is able to find a remote user that does not have the first transaction $\sim 60\%$  of the times, when half of the users in the network are assisting him. But If he wait a bit more (e.g. 30 more seconds), the first transaction will be spread in all the network, as shown in Figure \ref{fig:howMuchtimeSlower} (after 150 seconds $> 99\%$ have received the first transaction).

It is worth reminding that, in the examined scenarios, we present the cases when a mobile user can be cheated by a malicious one in terms of accepting the transaction. However, this does not mean that the malicious user managed to double spend some localcoins. In order to succeed in that, two users have to initiate block creations and succssfully verify both of these blocks. The necessity for high density comes from this need to allow the block creation procedure to both manage to spread in a big part of the network and detect the attempts for double spending.  

In summary, all three evaluation subsections are focused on describing the \textbf{area with high connectivity}, in which we argue that LocalCoin is applicable. This applicability depends on whether a normal transaction is properly spread to the whole network and a fake one is not spread and not verified. We examined both the case where the users are not moving and when they are moving either based on some available traces or by simulating their movement with the ONE simulator. In more detail, when users are static, we show that for a reasonable scenario, i.e., $|\mathcal{U}|=1000,r_{cov}=0.05$ the area consists of one major component implying feasibility of LocalCoin and impossibility of double spending attacks. When the users are walking, we show that with an average movement speed of 1kmph, they can complete their transactions in seconds while malicious users, even if they manage to find the proper time to initiate a fake transaction, a few minutes after that they will be detected in the block creation phase.

\section{Discussion}\label{sec:discussion}
We analysed LocalCoin by considering static graphs that are formed as random geometric graphs of which nodes are users' locations and the coverage area of the used wireless technology compared with the deployment area dictates whether two nodes are close enough to be connected. Static networks provide a lower bound in the performance of the protocol. In the case of mobile users, LocalCoin performs better because the transactions are more easily flooded to users and any malicious user has to not only consider the locations of the normal users but also their mobility, which is a very difficult task. An analysis of mobile users with different mobility patterns is part of our future work. It is notable that a conservative selection of the LocalCoin parameters (i.e., high values for $\mathit{mTr}, BS, \mathit{mVu}$ and $\mathit{aVd}$) can guarantee that double spending is not possible, however the performance of the protocol, in terms of time, will be hindered.

\subsection{Discussion about the assumptions of the analysis}\label{sec:assumptions}

The analysis presented in Section \ref{sec:analysis} is based on three major assumptions: 
\textbf{(A1)} The mobile users are uniformly distributed in the service area.
\textbf{(A2)} The mobile users can not hack their location.
\textbf{(A3)} The colluders of a malicious user are preselected and a malicious user is not able to bribe a normal user during the double spending attack.
Below we discuss the reasons that drove us to these three assumptions:

\textbf{(A1)}
The motivation behind distributing uniformly the mobile users in the examined area is that it makes the functionality of LocalCoin more challenging. Although it would be easier for a malicious user to identify critical locations to occupy if the users were not uniformly distributed, it would be also easier during the implementation of LocalCoin and the selection of its tuning parameters, to make such locations insufficient to create a virtual cut that can lead to a successful double spending attack. Figure~\ref{fig:basics} shows that a major connected component can be formed more easily in the case where the users are not uniformly distributed. 

\textbf{(A2)}
Although users' actual location is a core component of LocalCoin and a malicious user may consider asking a set of colluders to lie about their location and help him on double-spending, this is against his interest because the non-colluding neighbours of the colluders will immediately detect the attack and discard the block. Normal users know their location and the average coverage radius, so LocalCoin can handle such attacks by forcing each user to check the locations that her neighbours put during the block creation process. A parallel to the block verification mechanism can detect the location manipulation attempts and intercept the double spending attacks. 
Given that the actual locations are only needed for the calculation of the average distance between the verifiers of a block, the block creation and verification messages can be enlarged to contain the ids of the neighbours each verifier has when signing the block. Then the rule that dictates when a block can be created (i.e. when the average distance between the verifiers is higher than $aVd$) can be complemented by a second one that requires the number of the distinct users that are neighbours of the verifiers to be higher than a threshold. 
Although the incentives for each user to participate in such mechanism are not presented and analysed in detail, a simple defence mechanism that detects and punishes all the participants every time a location manipulation has been detected can be enough to create the necessary bias against the malicious users. 

\textbf{(A3)}
Our assumption is based on the fact that we envision LocalCoin to work on off-the-shelf mobile devices whose owners are using them on their needs while LocalCoin is running on the background. Mobile users are not aware of message exchange and the only way for a malicious user to double-spend is to implement another version of LocalCoin and install it in his colluders. 

\subsection{Getting assistance from a fixed network}
It is worth mentioning that LocalCoin can benefit from fixed networks in the small scale (university campus scale) to increase the speed of the message forwarding and decrease the importance of user's density. The access points of a fixed network can be treated as normal users who do not have storage capabilities (i.e. $\mathcal{T}_{AP} = \emptyset$, $\forall$ access point) and do not compete for the transaction fees but broadcast all the incoming messages to the other nodes of the fixed network who broadcast the messages to the associated mobile devices. Moreover, the access points can be used for the block creation process as a guarantee for the average distance between the mobile users. In more detail, each user who verifies the creation of a new block can attach a recently signed message that she received from the closest AP in order to provide a robust estimation of her location. However, these extra functionalities that are provided by a fixed network can only improve the performance of LocalCoin on the practical level. 

D2D architectures, with main representative being the 5G, are becoming more and more popular. The Wifi-direct technology is getting more mature and is adapted by many customer products, while, the design of LTE-direct is moving in this route. Motivated by advances in this direction and considering that any existing infrastructure, like an institutional network in a university campus, can only improve the coverage of LocalCoin, we argue that protocols like LocalCoin are applicable and implementable. 

\subsection{Power cost of LocalCoin}
LocalCoin, in contrast to Bitcoin, employs a defence mechanism against double spending that does not require powerful CPUs but requires an amount of messages to be exchanges between the mobile users. Based on measurements of the power consumption of the WiFi interface we estimate that a transaction with size 1 kilobyte consumes 0.00048/0.0004 joules when transmitted/received~\cite{7509388}. Although there exist studies that show that the power needs of WiFi is higher that the CPU~\cite{Carroll:2010:APC:1855840.1855861}, LocalCoin does not transfer messages 100\% of the time. A detailed analysis of the power consumption of LocalCoin is part of our future work. 

\section{Future Work}
In this work we analysed in detail the probability of a successful double spending, our next step is to analyse the proposed distributed blockchain and provide bounds regarding the required number of verifiers each users should have, the number of the blocks each user should store, on average, in order for the protocol to be functional and the relationship between the number of the stored blocks and earnings from the fees. 
Our future extensions of LocalCoin will be also focused on building defence mechanisms against other types of attacks where mobile users form coalitions and perform abnormally without their incentives being aligned with the incentives of the normal mobile users.

\section{Conclusion}
In conclusion, LocalCoin is a decentralised cryptocurrency built on principles similar to Bitcoin but avoids any need of the Internet and computing overheads. We used a distributed blockchain structure to store the produced transactions in the form of blocks and we proposed a set of four parameters to show the existing trade-offs in our proposal due to the mobile ad-hoc nature. These 4 parameters can be adjusted to provide the required security guarantees, in terms of double spending, and at the same time affect the maximum transaction rate.  LocalCoin is applicable in dense areas of mobile users and if the users are fully connected, the double spending is impossible. 
\bibliographystyle{abbrv}
\bibliography{sigproc} 



\begin{IEEEbiography}[{\includegraphics[width=1in,height=1.25in,clip,keepaspectratio]{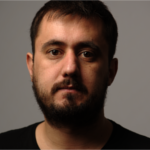}}]{Dimitris Chatzopoulos}
received his Diploma and his Msc in Computer Engineering and Communications from the Department of Electrical and Computer Engineering of University of Thessaly, Volos, Greece. He is currently a PhD student at the Department of Computer Science and Engineering of The Hong Kong University of Science and Technology and a member of HKUST-DT System and Media Lab. During the summer of 2014, he was a visiting PhD student at Ecole polytechnique federale de Lausanne (EPFL). His main research interests are in the areas of mobile computing, device--to--device ecosystems and cryptocurrencies.
\end{IEEEbiography}

\begin{IEEEbiography}[{\includegraphics[width=1in,height=1.25in,clip,keepaspectratio]{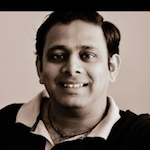}}]{Sujit Gujar}
is an Assistant Professor at the Machine Learning Laboratory@IIITH. Prior to this, he was a Sr. Research Associate at Indian Institute of Science. He worked as a post-doctoral researcher at Ecole polytechnique federale de Lausanne (EPFL). He also worked as a research scientist with Xerox Research Centre India where he contributed in developing a technology that enables enterprises to use crowdsourcing as a complimentary workforce. His research interests are Game Theory, Mechanism Design, Machine Learning, and Cryptography applied to modern web and AI applications such as Auctions, Internet  Advertising, Crowdsourcing,  and multi-agent systems. His doctoral thesis was awarded alumni medal for best doctoral thesis in the Department of Computer Science and Automation at Indian Institute of Science. He was a recipient of Infosys fellowship for his doctoral research. He has co-authored 5 journal publications, 1 book chapter and 22 conference/workshop papers. He has 11 patents on his name.
\end{IEEEbiography}


\begin{IEEEbiography}[{\includegraphics[width=1in,height=1.25in,clip,keepaspectratio]{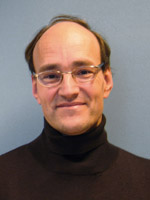}}]{Boi Faltings}
 is a full professor of computer science at the Ecole Polytechnique Federale de Lausanne (EPFL),  where he heads the Artificial Intelligence Laboratory, and has held visiting positions at NEC Research Institute, Stanford University and the HongKong University of Science and Technology. He has co-founded 6 companies in e-commerce and computer security and acted as advisor to several other companies. Prof. Faltings has published over 300 refereed papers and graduated over 30 Ph.D. students, several of which have won national and international awards. He is a fellow of the European Coordinating Committee for Artificial Intelligence and a fellow of the Association for Advancement of Artificial Intelligence (AAAI). He holds a Diploma from ETH Zurich and a Ph.D. from the University of Illinois at Urbana-Champaign.
\end{IEEEbiography}


\begin{IEEEbiography}[{\includegraphics[width=1in,height=1.25in,clip,keepaspectratio]{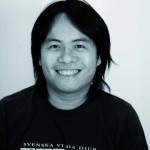}}]{Pan Hui}
received his Ph.D degree from Computer Laboratory, University of Cambridge, and earned his MPhil and BEng both from the Department of Electrical and Electronic Engineering, University of Hong Kong. He is currently a faculty member of the Department of Computer Science and Engineering at the Hong Kong University of Science and Technology where he directs the HKUST-DT System and Media Lab. He also serves as a Distinguished Scientist of Telekom Innovation Laboratories (T-labs) Germany and an adjunct Professor of social computing and networking at Aalto University Finland. Before returning to Hong Kong, he has spent several years in T-labs and Intel Research Cambridge. He has published more than 150 research papers and has some granted and pending European patents. He has founded and chaired several IEEE/ACM conferences/workshops, and has been serving on the organising and technical program committee of numerous international conferences and workshops including ACM SIGCOMM, IEEE Infocom, ICNP, SECON, MASS, Globecom, WCNC, ITC, ICWSM and WWW. He is an associate editor for IEEE Transactions on Mobile Computing and IEEE Transactions on Cloud Computing.
\end{IEEEbiography}

\end{document}